\documentclass[preprint,review,12pt,authoryear]{elsarticle}

\usepackage{amssymb}
\usepackage{amsmath}
\usepackage[gen]{eurosym}
\usepackage{url}
\usepackage{setspace}
\usepackage{booktabs}
\usepackage{array}
\usepackage{siunitx}
\usepackage{pdflscape}
\usepackage{makecell}
\usepackage{tabularx}
\usepackage{graphicx}
\usepackage[utf8]{inputenc}
\usepackage{subcaption}
\usepackage{grffile} % Allows spaces and special characters in filenames
\usepackage[margin=1in]{geometry} % Adjust margins to fit more on a page
\usepackage{multirow}
\usepackage[hidelinks]{hyperref}
\usepackage{placeins}
\journal{Energy Economics}

\begin{document}
%TC:ignore 
\begin{frontmatter}

%Previous
\title{Disentangling the Impact of Contracts for Difference on Sequential Electricity Markets: An Iberian Case Study} 
%Proposed
%\title{Contracts for Difference in Sequential Electricity Markets: Policy Design Implications from the Iberian Market}
%% Article title

\author[1,3,4]{Silke Johanndeiter}%\corref{cor1}}
\affiliation[1]{organization={Ruhr-Universität Bochum, Germany}}
%,
   % addressline={Universitätsstr. 150}, 
   % city={44801 Bochum},
   % country={Germany}}
\affiliation[3]{organization={University of Cologne, Germany}}
\affiliation[4]{organization={Institute of Energy Economics at the University of Cologne (EWI), Germany}}   

\author[2]{Ricardo Faia}
\affiliation[2]{organization={Research Group on Intelligent Engineering and Computing for Advanced Innovation and Development (GECAD), LASI, ISEP, Polytechnic of Porto (IPP), Portugal}}
%,   addressline={R. Dr. António Bernardino de Almeida, 431},
  %  postcode={4249-015}, 
   % city={Porto},
   % country={Portugal}}
\author[2]{Fernando Lezama
%\corref{cor1}
}
\cortext[cor1]{Corresponding author E-mail address: johanndeiter@wiso.uni-koeln.de}
%\cortext[cor1]{Corresponding author E-mail address: flz@isep.ipp.pt}
\author[2]{Zita Vale}

%Abstract
\begin{singlespace}
\begin{abstract}
Two-way Contracts for Difference (CfDs) are now the mandatory design for governmental renewable support schemes in the European Union. They specify payments based on the difference between a strike and a reference price. While they mitigate market risks, production-based CfDs are known to distort short-term markets. We quantify these distortions for CfD designs that vary in strike and reference price in a sequential electricity market. We theoretically derive optimal day-ahead and intraday bidding strategies for CfD-supported wind generators. To disentangle the effects of day-ahead and intraday strategies on aggregate market outcomes, we implement them in a sequential market simulator calibrated to historical bids from the Iberian electricity market. Across all considered CfD designs, we find that wind curtailment is driven primarily by day-ahead market incentives, whereas prices change in up to 45\% of hours in both the intraday and day-ahead markets. Although CfDs systematically alter wind offer and bid prices in intraday markets, price changes mainly result from horizontal shifts of intraday supply and demand curves caused by preceding distortions of day-ahead market sales. Compared to a baseline without a production-based CfD, these shifts raise the frequency of negative intraday prices, while increasing intraday prices by up to 93\,\euro{}/MWh in other hours. Balancing providers are particularly sensitive to these intraday price changes. Changing the reference price period redistributes rather than eliminates operational distortions across sequential markets. Where preserving short-term market signals is a policy objective, the findings support settlement structures that decouple CfD payments from realised generation.
\end{abstract}

%Overall, our results provide quantitative evidence of intraday in addition to widely studied day-ahead market distortions induced by production-based CfDs, supporting the case for production-independent CfD payment designs.


\begin{highlights}
%Previous
%\item Quantification of Contracts for Difference-induced distortions in sequential markets
% \item Derivation of CfD-dependent bidding strategies in day-ahead and intraday auctions. 
% \item A sequential market simulator identifies market distortions in an Iberian case study.
% \item Day-ahead bidding drives wind curtailment, which is partly offset by intraday sales.
% \item CfDs alter intraday prices in up to 45\% of hours, at least as often as day-ahead prices.
% \item All CfD designs and strike prices increase the incidence of negative intraday prices.
%Proposed
\item CfD reference-price design redistributes distortions across sequential markets.
\item Day-ahead incentives drive wind curtailment, partly offset by intraday trading.
\item CfDs alter intraday prices in up to 45\% of hours, at least as often as day-ahead prices.
\item Strike and reference prices jointly determine curtailment and price distortions.
\item Production-independent settlement better preserves short-term market signals.

\end{highlights}

%% Keywords
\end{singlespace}
\end{frontmatter}


%TC:endignore 
\newpage
\section{Introduction}
\label{sec1:intro}
Contracts for Difference (CfDs) are widely used to mitigate price risk for renewable electricity producers. They stipulate payments between contract parties based on the difference between a reference market price and a predetermined strike price. Since 2024, two-way CfDs are the mandatory design for governmental support schemes of renewables in the European Union (EU) \citep{eu2024}. Two-way indicates that they allow for payments from and to both contract parties. Therefore, governmental CfDs constitute a hedging instrument for renewable generators as well as the government, which can redistribute CfD payments to consumers. Increasingly, CfDs are also used as private hedging instruments for renewables, particularly power purchase agreements (PPAs) \citep{beiter2024, lezama2025evaluating}. Two-way CfDs can be designed in various ways, differing in their settlement unit and reference price \citep{johanndeiter2025}. The literature suggests that the design of CfDs not only affects investment choices but also the behaviour of renewables in short-term electricity markets, with important implications for the efficiency of electricity markets. However, quantitative analyses remain largely limited to the effect on the day-ahead (DA) market, with only a few studies examining the impact on subsequent auctions such as intraday (ID) auctions, which we discuss in detail in Section \ref{sec:lit}. 

This work provides a quantitative assessment of the impact of different types of CfDs on short-term electricity markets, taking into account their sequential nature. First, we theoretically derive optimal spot market bidding strategies of renewables that are subject to different types of CfDs (Section \ref{sec:theory}). They participate in a competitive uniform-price DA auction that is followed by an ID auction. Second, we implement the derived bidding strategies in a simulation framework that matches supply and demand curves in a sequential electricity market (Section \ref{sec:method}). We apply our theory and our framework to historically observed bids submitted by wind farms in the Iberian electricity market (MIBEL). Finally, we disentangle the impact of bidding strategies on aggregate DA and ID market outcomes, including electricity prices and wind curtailment (Section \ref{sec:results}). By comparing different strike price levels and reference-price periods, the analysis identifies how CfD design choices redistribute operational distortions across sequential markets. These results provide evidence for policymakers when assessing whether CfD settlements should remain linked to realised generation and how alternative reference-price designs affect short-term market efficiency. Section \ref{sec:conclusion} concludes and discusses the resulting policy implications.

\section{Literature review}
\label{sec:lit}
Two-way CfDs have two main design elements that determine their impact on investment and operational decisions: the \textit{reference price}, whose difference from the strike price determines payments between the contract parties, and the \textit{unit of payment}, namely, electricity generated (in MWh) or capacity installed (in MW). The effect of these design elements on investment decisions, such as the choice of location, rotor size, or hub height of wind farms, materialises in the strike price. This price is typically determined through competitive auctions, in which projects offering the lowest strike prices are awarded CfDs \citep{anatolitis2017}. The impact of these design elements on operational decisions, in turn, is reflected in the bids submitted to short-term electricity markets.

%\textbf{Investment incentives.}
\subsection{Investment incentives}
Investment incentives of conventional CfDs are determined by the reference price, which is defined as the individual market revenues of renewable generators in (quarter-)hourly trading windows. Consequently, power plants subject to conventional CfDs receive the strike price for each MWh generated or MW of capacity installed, thereby fully hedging price risk \citep{wagner2026}. The reference price of a production-based conventional CfD is the (quarter-)hourly DA market price, causing the scheme to resemble a feed-in tariff (FIT). In this case, power plants with higher full load hours are more likely to obtain a CfD because they can bid lower strike prices in a competitive auction \citep{johanndeiter2025}. Capacity-based support schemes, in turn, have been criticised for giving rise to the `steel-in-the-ground' phenomenon, i.e., incentivising wind turbine designs that maximise installed capacity rather than electricity generation \citep{boute2012,meus2021}. Both issues can be addressed by more sophisticated CfD designs that decouple individual revenues from CfD payments. Instead of comparing individual revenues with the strike price, these schemes use a reference price based on the revenues of a reference power plant. The output of this reference plant may be determined by a location-specific benchmark \citep{newbery2023}, derived from weather data, or based on the average generation profile within a bidding zone \citep{huntington2017,schlecht2024}. Such a design enables power plants that outperform the reference to generate additional revenues, allowing them to submit lower strike prices \citep{johanndeiter2025}.
For production-based CfDs, lower strike prices can be submitted by power plants with above-average market values, i.e.\,generation-weighted market revenues, which are associated with lower system costs \citep{ueckerdt2013}. Conversely, capacity-based CfDs foster investment with capacity-weighted revenues above the reference that do not necessarily coincide with higher market values \citep{johanndeiter2025b}. Therefore, compared to production-based CfDs, they induce different plant designs and locations. Previous work based on wind investment models \citep{schmidt2013,may2017,meus2021} and historic data for Germany \citep{grothe2013} shows that support scheme design indeed affects location and design of wind farms. However, this incentive only exists when the reference power plant's revenues differ sufficiently from the investor's own expected revenues. In turn, this implies higher revenue risks \citep{johanndeiter2025b} that can increase investors' capital costs \citep{dukan2025}. Furthermore, for CfDs used as private medium-term hedging instruments, \cite{lezama2025evaluating} show how the contract structure and strike price redistribute price risk between contracting parties. 

%\textbf{Operational incentives.}
\subsection{Operational incentives}
The fact that production-based conventional CfDs effectively resemble a FIT also shapes their operational incentives: They incentivise maximising output during the operational phase, regardless of market prices. This effect is called ``produce-and-forget'' incentive and is shown by \cite{andor2016} in a theoretical model. \cite{devos2015} and \cite{frondel2022} provide empirical evidence that this operational strategy contributes to the occurrence of negative market prices. Considering operational constraints in a deterministic partial equilibrium model of the Danish electricity market, \cite{rosnes2014}' results add that the strategy also leads to inefficient operation of thermal power plants, particularly in terms of the number of costly start-ups, increasing emissions and system costs. The occurrence of negative prices can be reduced by restricting CfD payments to hours with non-negative prices as shown in a partial equilibrium model of a power market including thirteen connected countries \citep{veenstra2024}. Production-based CfDs with reference prices equal to average market values, in turn, are known to distort operational decisions differently: they add a constant negative or positive premium to market revenues, constituting virtual revenues or costs equal to anticipated CfD payments. Virtual costs cause power plants to withhold dispatch below a price threshold, while virtual revenues induce dispatch even at negative prices. \cite{frondel2022}'s empirical results demonstrate that a market premium (MP), comparable to this CfD design, reduces the occurrence of negative prices compared to a FIT. However, model-based analyses demonstrate that it also induces negative prices up to the level of the premium, as shown by a stylised economic model of the German electricity market \citep{pahle2016}, agent-based simulations \citep{winkler2016} and \cite{frey2020}, and optimisation models of the European energy system \citep{veenstra2024,johanndeiter2025}. Therefore, renewable generators are dispatched even when market prices fall below their technical variable costs, particularly down to the negative value of the premium, thereby reducing curtailment. This result is consistent with a unit commitment model of the Spanish market \citep{chaves2017}, an optimisation model of the European market \citep{ozdemir2020} and an empirical analysis of wind offshore power plants in Great Britain \citep{Steenberghe2026}. Virtual costs also affect the ID market and balancing payments; when the reference price exceeds the strike price, generators must pay the difference to the counterparty for every MWh generated. As a result, they can increase profits by curtailing output and purchasing DA committed volumes in the ID market, or by settling them at the imbalance price if the respective prices fall below the payment obligation \citep{schlecht2024,khodadadi2024}. Based on historical electricity price data from the German DA and ID market, \citet{stecher2025} quantify the number of hours where such incentives are prevalent, finding them most relevant in high-price years such as 2022, affecting 23\% of hours. However, historical evidence from Germany, France, and Belgium indicates that negative ID prices were correlated with forecast errors and imbalance settlement prices, suggesting no historical relationship with support scheme design \citep{devos2015}. More recent evidence based on panel data from offshore wind farms in Great Britain indicates that generators are less likely to curtail output when CfD payments exceed the imbalance price, and more likely to do so when the opposite applies \citep{Steenberghe2026}.
Due to these operational distortions induced by production-based CfDs, several scholars argue for the introduction of production-independent CfDs \citep{huntington2017,newbery2023,schlecht2024}.  
Table \ref{tab:lit} summarises previous work on the effect of renewable support schemes on operational decisions and market prices.

\begin{table}[h!]
    \centering
    \footnotesize
    \caption{Overview of previous literature on operational effects induced by renewable support schemes}
                \label{tab:lit}
      \begin{tabular}{>{\raggedright\arraybackslash}p{3cm} p{5cm} p{5cm}}
        \toprule
        Paper & Method & Result \\
        \midrule
        \cite{rosnes2014} & Deterministic partial equilibrium model of Danish electricity market & Operational inefficiencies highest under FIT, lower for MP, lowest for capacity-based support scheme \\
        \cite{devos2015} & Empirical analysis of historic German, French and Belgian electricity market results & More negative prices in Germany due to FIT \\
        \cite{andor2016} & Theoretical model & Dispatch at any price under FIT \\
        \cite{pahle2016} & Stylised economic model of German electricity market & Negative prices caused by MP; lower curtailment \\
        \cite{winkler2016} & Agent-based simulation of the German electricity market & Negative prices caused by MP \\
        \cite{frey2020} & Agent-based simulation of the German electricity market & Negative prices caused by MP \\
        \cite{ozdemir2020} & Optimisation model of the European energy system, support schemes implemented via constraints & Production-based support reduces curtailment compared to capacity-based support \\
        \cite{frondel2022} & Nonparametric modeling using historic German data & Reduction of negative prices for MP compared to FIT \\
        \cite{schlecht2024} & Conceptual discussion & Production-based CfDs distort DA and ID market bids \\
        \cite{khodadadi2024} & Conceptual discussion & Production-based CfDs distort DA and ID market bids \\
        \cite{veenstra2024} & Partial equilibrium model of European power market & CfDs that restrict payments to hours with non-negative prices reduce occurrence of negative prices \\
        \cite{johanndeiter2025} & Optimisation model of the European energy system, support schemes implemented by altering cost parameters & Production-based CfDs distort market prices and curtailment \\
         \cite{Steenberghe2026} & Panel data analysis of offshore wind farms subject to different support schemes in Great Britain & Production-based CfDs decrease DA curtailment and distort response to imbalance prices \\
        \bottomrule
    \end{tabular}
     \begin{minipage}[c]{\textwidth}
     {\scriptsize
			\addtolength{\footnotesep}{2pt}
			\noindent
     Acronyms: FIT = Feed-in Tariff, MP = Market Premia}.
\end{minipage}
   % \label{tab:literature}

\end{table}

%\textbf{Research gap.}
\subsection{Research gap}
Overall, a production-independent CfD design with difference payments based on revenues of a reference power plant appears favourable from an operational perspective, but its induced investment incentives remain ambiguous. While effects of investment incentives have been investigated widely, a detailed quantitative assessment of operational distortions is needed to fully evaluate this trade-off. Due to the recent introduction of two-way CfDs as a mandatory supports scheme in the European Union \citep{eu2024}, several member states are reforming their national support schemes. These reforms require policy choices regarding whether CfD settlements should depend on realised generation, how reference prices should be defined, and how strike-price levels interact with short-term market incentives. Previous quantitative studies on operational effects remain largely limited to the DA market, while impacts on ID markets and balancing behaviour have been addressed mainly conceptually and quantified only to a limited extent. This study addresses this gap by quantifying how alternative CfD designs redistribute distortions across sequential DA and ID markets, providing evidence to support these policy-design choices. To the best of our knowledge, this study is the first to theoretically derive CfD-dependent bidding strategies in a sequential electricity market and to quantitatively disentangle their impact on aggregate market outcomes.
%This work intends to provide scientific evidence for policy-maker to make an informed choice.

\section{Theoretical Background}
\label{sec:theory}
We consider a renewable power plant participating in an electricity market that consists of two sequential uniform-price-auctions for electricity taking place one day before delivery (i.e., DA market) and a time closer to delivery (i.e., ID market).
We assume the renewable power plants to be subject to private or governmental two-way CfDs that stipulate payments between the contract parties based on an ex ante defined strike price and an ex post reference market price. The definition of the reference market price differs by CfD types that are additionally distinguished by the unit of payment (volume of electricity generated or capacity installed). In the following, we derive optimal bidding strategies for renewable power plants in the DA and ID auctions, depending on the specific design of the CfD. Particularly, we consider renewable power plants' expected short-term profit functions at the time of each auction, when market clearing prices and available capacity are unknown, and investment costs are sunk. 

\subsection{Market setup}
\label{sec:theory_market}
A power plant in our market setup has the following ex post short-term profit function in a given hour $t$:
\begin{align}
\label{eq:profit_expost}
  \Pi_t^\text{type}=q_t^{DA} \cdot p_t^{DA} 
  + q_t^{ID} \cdot p_t^{ID} 
  -  Q_t \cdot c  +  (Q_t -q_t^{DA}-q_t^{ID}) \cdot p_t^{imb} + P^\text{type},
\end{align} 
where, $q^{DA}_t$ represents the quantity sold day ahead and $q^{ID}_t$ the quantity sold (bought) ID if positive (negative). Variable $Q_t$ is defined as total actual generation. The variables $p^{DA}_t$ and $p^{ID}_t$ denote the clearing prices of the DA and ID auction, respectively, while $c$ describes constant marginal costs of electricity generation. Differences between actual generation, $Q_t$, and scheduled generation, $q^{DA}_t + q^{ID}_t$, are penalised with the imbalance price $p_t^{imb}$, whose sign depends on the system balance. If system load is higher than system generation (i.e.\ the grid frequency is lower than 50\,Hz), $p_t^{imb}$ is positive and negative otherwise. We assume that the power plant's imbalance is aligned with the overall system imbalance, such that $(Q_t -q_t^{DA}-q_t^{ID}) \cdot p_t^{imb} \leq 0$. Therefore, power plants would curtail output if available capacity is higher than total wholesale market sales in order to avoid paying the balancing price for the excess generation. Furthermore, we assume total sales not to exceed available generation, $q_t^{DA}+ q_t^{ID} \leq Q_t$, and that generators are not allowed to purchase electricity ID if their net generation reaches zero, $q_t^{DA}+ q_t^{ID} \geq 0$.
Hence, all renewable power plants that successfully settle are net sellers over all market sequences. Overall, these assumptions imply that $q_t^{DA}+ q_t^{ID} \in [0,\,Q_t]$. Finally, power plants can be subject to CfD payments, $P^\text{type}$, defined by the difference of strike price $S$ and a possibly time-dependent reference price $p_{t}^{R,\text{type}}$ that differ by CfD type $\in [\text{none},\text{cap},\text{hourly},\text{constant}]$. Besides a case with no CfD ($P^\text{none}=0$), we consider the following CfD types, whose specific payment functions $P^\text{type}$ are presented in Section \ref{sec:bidding}:
\begin{itemize}
    \item \textit{cap}: the reference price $p^\text{R,cap}_{t} = \bar r$ is constant throughout the entire contract duration $T$, and its difference to the strike price is paid by capacity $\bar Q$,
    \item \textit{hourly}: the reference price is defined as the hourly DA market price $p_t^{R,\text{hourly}}=p_t^{DA}\,\forall \,t\in T$ and its differences to the strike price are paid by total production $Q_t$,
    \item \textit{constant}: the reference price is defined as a constant value $p_t^{R,\text{constant}}=v_{\tau}\,\forall \,t\in \tau$ over a certain period $\tau$ and its differences to the strike price are paid by total production $Q_t$.
\end{itemize}

In this market setup, power plants submit supply schedules to auctions $s\in\{\text{DA},\text{ID}\}$ and, in the ID market, may also submit demand schedules to buy electricity. Each schedule consists of up to $N$ steps indexed by $n$, with pairs $(q_n,p_n)$, where $q_n$ is the quantity offered or bid at price $p_n$. For a uniform market clearing price $p_t^s$, supply steps clear when their offer price is below the clearing price, while ID demand steps clear when their bid price is greater than or equal to the ID clearing price:
\begin{align}
S_t^s(p_t^s;p_n) &=
\begin{cases}
q_n, & \text{if } p_n < p_t^s,\\
0, & \text{otherwise},
\end{cases}
&
D_t^{ID}(p_t^{ID};p_n) &=
\begin{cases}
-q_n, & \text{if } p_n \geq p_t^{ID},\\
0, & \text{otherwise}.
\end{cases}
\end{align}
Hence, $q_t^s=\sum_{n=1}^{N}S_t^s(p_t^s;p_n)$ and the net ID position is $q_t^{ID}=\sum_{n=1}^{N}S_t^{ID}(p_t^{ID};p_n)+\sum_{n=1}^{N}D_t^{ID}(p_t^{ID};p_n)$, where purchases enter with a negative sign. In the following, we use $(q_n,p_n)^{S,DA}$ for the $n$th DA offer, $(q_n,p_n)^{S,ID}$ for the $n$th ID offer, and $(q_n,p_n)^{D,ID}$ for the $n$th ID bid.

%----

\subsection{Bidding strategies}
\label{sec:bidding}
We derive optimal bidding strategies of renewable power plants in the described market setting and that are subject to (i) no production-based CfD, (ii) the production-based hourly CfD and (iii) the production-based constant CfD.

Throughout the derivation, subscript $t$ denotes the delivery hour and $n$ the step of an offer or bid schedule. Superscripts DA and ID denote the market sequence, while $S$ and $D$ distinguish supply offers from demand bids. Variables with a hat denote expectations formed at the time of the auction, before market clearing and before final available generation is known. Variables without a hat denote realised or already-cleared values. Thus, $\hat p_t^s$ and $\hat q_t^s$ are expected clearing prices and quantities, while $p_t^s$ and $q_t^s$ denote their realised counterparts.

\subsubsection{No production-based CfD}
\label{sec:bid_nocfd}

Without a production-based CfD, the DA bidding strategy follows directly from expected short-term profits,
\begin{align}
 \hat \Pi^{DA,\text{none}}_t
 =
 \hat q_t^{DA}\cdot(\hat p_t^{DA}-c).
\end{align}
Profits increase with sold quantity whenever the DA clearing price exceeds marginal generation costs. The optimal DA offer is therefore to submit the available forecasted generation at marginal cost, $(q_1,p_1)^{S,DA}=(\hat Q_t^{DA},c)$.

After the DA auction, the plant updates its available generation forecast to $\hat Q_t^{ID}$ and submits ID offers or bids according to
\begin{align}
\label{eq:profits_id_none}
\hat \Pi_t^{ID,\text{none}}
=
\hat q_t^{ID}\cdot \hat p_t^{ID}
+
q_t^{DA}\cdot p_t^{DA}
-
(q_t^{DA}+\hat q_t^{ID})\cdot c
+
\min\{0,(\hat Q_t^{ID}-q_t^{DA}-\hat q_t^{ID})\cdot \hat p_t^{imb}\}.
\end{align}
If $\hat Q_t^{ID}\geq q_t^{DA}$, additional ID sales affect generation but not imbalance payments. The plant therefore offers its remaining available generation at marginal cost, $(q_1,p_1)^{S,ID}=(\hat Q_t^{ID}-q_t^{DA},c)$, and may buy back its DA commitment whenever the ID price is below marginal generation costs, $(q_1,p_1)^{D,ID}=(-q_t^{DA},c)$. If $\hat Q_t^{ID}<q_t^{DA}$, the plant first bids for the missing generation at the expected imbalance price, $(q_1,p_1)^{D,ID}=(\hat Q_t^{ID}-q_t^{DA},\hat p_t^{imb})$, and may additionally buy back the remaining DA commitment if the ID price is below marginal generation costs, $(q_2,p_2)^{D,ID}=(-\hat Q_t^{ID},c)$. The detailed case derivations are provided in~\ref{sec:derivation}.

Overall, without a production-based CfD, quantity bids are driven by expected generation, while price bids reflect marginal generation costs and avoided imbalance payments. These bids can therefore be considered efficient in the absence of CfD-induced virtual costs or revenues (cf. Section \ref{sec:lit}).

\paragraph{Capacity-based CfD}
It is straightforward to show that the same bidding strategies hold for any CfD, where payments are not defined based on actual generation. For the capacity-based CfD defined in Section \ref{sec:theory_market}, ex post profits are given by
\begin{align}
\label{eq:profit_expost_finCfD}
  \Pi^\text{cap}_t= q_t^{DA} \cdot p_t^{DA} +q_t^{ID} \cdot p_t^{ID}
  - Q_t  \cdot c + (Q_t -q_t^{DA}-q_t^{ID}) \cdot p_t^{imb} + \bar Q  \cdot (S - \bar r)  .
\end{align}
For all derivations above, the term $\bar Q  \cdot (S - \bar r)$ that describes the CfD payments based on installed capacity $\bar Q$, would be eliminated, as DA and ID sales/purchases only affect generation and not the installed capacity. Hence, bids in both auctions are the same as in the absence of any CfD (cf. also \cite{khodadadi2024}).

\subsubsection{Production-based CfD with hourly reference price}
\label{sec:bid_basiccfd}
Now, we consider the production-based \textit{hourly} CfD, with the reference price defined as the DA market price in time step $t$, $p^{R,\text{hourly}}=p_t^{DA}$, and payments made by volume produced. For this CfD, the ex post profit function is given by
\begin{align}
\label{eq:profit_expost_basicCfD}
  \Pi^\text{hourly}_t= q_t^{DA} \cdot p_t^{DA} +q_t^{ID} \cdot p_t^{ID}
  - Q_t  \cdot c +  (Q_t -q_t^{DA}-q_t^{ID}) \cdot p_t^{imb} + Q_t  \cdot (S - p_t^{DA}). 
\end{align}
For a power plant subject to the hourly CfD, profits from selling offer $(q_n,p_n)^{S,DA}$ DA are at least zero, if $
\hat  \Pi^{DA,\text{hourly}}_t= q_n \cdot p_n
  + q_n  \cdot (S - p_n)
  - q_n  \cdot c = q_n  \cdot (S - c) \geq 0  \Leftrightarrow  S \geq c $.
The inequality shows that profits are increasing for any $q_n$, if the strike price is higher than the marginal costs of electricity generation. Effectively, the DA price is irrelevant for the power plant as it receives $(S - c)$ for each kWh sold DA. Hence, under the derived condition, power plants sell available power generation $\hat Q^{DA}_t$ for any price up to the minimum price, $p^{DA}_{min}$, such that the 1$^{st}$ and only offer is $(q_1,p_1)^{S,DA}=(\hat Q^{DA}_t,p^{DA}_{min})$ if $\ S\geq c$.

The bid implies that renewables subject to an hourly CfD sell power, even if the market price does not cover the marginal cost of electricity generation. This result is in line with previous literature on the CfD-induced distortion of the DA market, known as the `produce-and-forget' effect (cf. Section \ref{sec:lit}).

After the DA auction is cleared, expected ID profits under the hourly CfD become:
\begin{align}
\label{eq:profits_id_bcfd}
\hat \Pi_t^{ID,\text{hourly}} = \hat q_t^{ID}\cdot \hat p_t^{ID}
+ q_t^{DA}\cdot p_t^{DA} - (q_t^{DA}+\hat q_t^{ID})\cdot(c-S+p_t^{DA}) + \notag\\ 
\min\{0,\,(\hat Q_t^{ID}-q_t^{DA}-\hat q_t^{ID})\cdot \hat p_t^{imb}\}.
\end{align}
Compared to the case without a production-based CfD, marginal generation costs are replaced by the virtual marginal cost $c-S+p_t^{DA}$ (c.f Section \ref{sec:lit}). Therefore, if $\hat Q_t^{ID}\geq q_t^{DA}$, the optimal ID offer is $(q_1,p_1)^{S,ID}=(\hat Q_t^{ID}-q_t^{DA},c-S+p_t^{DA})$, while the corresponding buy-back bid is $(q_1,p_1)^{D,ID}=(-q_t^{DA},c-S+p_t^{DA})$. If $\hat Q_t^{ID}<q_t^{DA}$, the first demand bid for missing generation remains unchanged, $(q_1,p_1)^{D,ID}=(\hat Q_t^{ID}-q_t^{DA},\hat p_t^{imb})$, because CfD payments do not affect imbalance payments. The additional buy-back bid is then $(q_2,p_2)^{D,ID}=(-\hat Q_t^{ID},c-S+p_t^{DA})$. The detailed price-threshold derivations are provided in~\ref{sec:derivation}.

Thus, hourly CfD payments enter ID offers and buy-back bids as virtual marginal costs or revenues whenever the corresponding quantity affects physical generation. If $p_t^{DA}>S$, the plant must pay the CfD counterparty for each MWh generated, increasing its incentive to buy back DA commitments and curtail output. If $p_t^{DA}<S$, positive CfD payments reduce this incentive and may require negative ID prices for buy-back to become profitable.
%------------

\subsubsection{Production-based CfD with constant reference price}
Finally, we consider a CfD with a reference price that is constant over a certain period of time, $\tau$, $p^{R,\text{constant}}=v_{\tau} \, \forall \, t \in \tau$. For a power plant subject to this type of CfD, which we deem \textit{constant} CfD, ex post profits become:
\begin{align}
\label{eq:profit_expost_2wayCfD}
  \Pi_t^\text{constant}= q_t^{DA} \cdot p_t^{DA} +q_t^{ID} \cdot p_t^{ID}
  + (Q_t -q_t^{DA}-q_t^{ID}) \cdot p_t^{imb} + Q_t  \cdot (S - v_{\tau} - c) 
\end{align}
Hence, in the DA market, bid $(q_1,p_1)^{S,DA}$ with $q_1=\hat Q_t^{DA}$ is more or equally profitable than selling nothing DA, if $
  \Pi^{DA,\text{constant}}_t=\hat q_t^{DA} \cdot \hat p_t^{DA} + \hat q_t^{DA} \cdot (S - \hat v_{\tau} - c) \geq 0 \notag
  \Leftrightarrow  \hat p_t^{DA} \geq  (c  - S + \hat v_{\tau})$.
This implies that the power plants offer available power $q_1=\hat Q_t^{DA}$ for a price that is equal to its marginal cost of electricity generation plus anticipated payments that -- depending on expectations on the reference price, $\hat v_{\tau} $, and the strike price $ S$ -- can be positive or negative $p_1=c + \hat v_{\tau} - S$, i.e.\ $(q_1,p_1)^{S,DA}=(\hat Q_t^{DA},c + \hat v_{\tau} - S)^{S,DA}$. 

In contrast to the hourly CfD, the constant production-based CfD does not incentivise generating electricity at any price. However, depending on the relationship between the strike price and the reference price, it can lead to power plants dispatching at prices below or above technical marginal costs of electricity generation, causing inefficient dispatch.

After the DA auction is cleared, the expected ID profits under the constant production-and-forget incentive CfD become:
\begin{align}
\label{eq:profits_id_2waycfd}
\hat \Pi_t^{ID,\text{constant}} = \hat q_t^{ID} \cdot \hat p_t^{ID} + q_t^{DA} \cdot p_t^{DA}
  -  (q_t^{DA}+\hat q_t^{ID}) \cdot (c\ -  S + \hat v_{\tau}) \notag \\   +  \min \{0,\,(\hat Q_t^{ID}-q_t^{DA}-\hat q_t^{ID})\cdot \hat p_t^{imb}\} 
\end{align}

%-----------------
% It becomes evident that profits are equivalent to the hourly CfD (cf. Section \ref{sec:bid_basiccfd}), when exchanging the reference price in Equation (\ref{eq:profits_id_bcfd}) for $v_{\tau}$. Therefore, the bidding logic is also equivalent:
% \begin{itemize}
%     \item For $\hat Q_t^{ID}\geq q_t^{DA}$, $(q_1,p_1)^{S,ID}=(\hat Q_t^{ID}-q_t^{DA},c - S + \hat v_{\tau})$ is the optimal 1$^{st}$ offer.
%     \item For $\hat Q_t^{ID}< q_t^{DA}$, $(q_1,p_1)^{D,ID}=(\hat Q_t^{ID}-q_t^{DA},\hat p_t^{imb})$ is the optimal 1$^{st}$ bid. 
%     \item For both cases, $(q_2,p_2)^{D,ID}=(\max\{-q_t^{DA},-\hat Q_t^{ID}\},c - S + \hat v_{\tau})$ is the optimal $2^{nd}$ bid.
% \end{itemize}
%---------
It becomes evident that profits are equivalent to the hourly CfD case after replacing $p_t^{DA}$ in Equation~(\ref{eq:profits_id_bcfd}) by $\hat v_{\tau}$. Therefore, the bidding logic is also equivalent: for $\hat Q_t^{ID}\geq q_t^{DA}$, the optimal ID offer is $(q_1,p_1)^{S,ID}=(\hat Q_t^{ID}-q_t^{DA},c-S+\hat v_{\tau})$; for $\hat Q_t^{ID}<q_t^{DA}$, the missing-generation bid is $(q_1,p_1)^{D,ID}=(\hat Q_t^{ID}-q_t^{DA},\hat p_t^{imb})$; and in both cases the buy-back bid is $(q_2,p_2)^{D,ID}=(\max\{-q_t^{DA},-\hat Q_t^{ID}\},c-S+\hat v_{\tau})$.

Analogously to the hourly production-based CfD, the constant CfD induces virtual marginal costs at the level of anticipated CfD payments that distort ID price bids when the corresponding quantity bid affects electricity generation.
While strategies are equivalent, it is important to note that under the constant production-based CfD, price bids are constant over the reference period used to calculate $v_{\tau}$. In contrast, under the hourly CfD, they vary from hour to hour with the DA market price, since the reference price equals the market price by definition, limiting the average deviation between the two. Effects on price bids and resulting curtailment under the constant CfD are therefore expected to be more pronounced.

\subsection{Summary}
Table \ref{tab:theory_summary} sums up the identified bidding strategies. Overall, renewable power plants submit offer and bid schedules for four different reasons: (i) to offer their \textbf{available capacity} DA; (ii) to offer their \textbf{remaining generation} ID if the ID forecast, $\hat Q_t^{ID}$, exceeds DA sales, $q_t^{DA}$; (iii) to buy \textbf{missing generation}, if capacity forecasts decline below DA commitments ($\hat Q_t^{ID} < q_t^{DA}$) and (iv) to \textbf{buy back} DA commitments ID to save production costs and if relevant, CfD payments. While the strategy to buy missing generation is not affected by production-based CfDs, price bids in all other situations reflect virtual marginal costs arising from CfD payments. An exception is the DA market price bid under the hourly CfD. In this case, the reference price cancels out of the profit function, such that the power plant bids at the auction’s minimum price regardless of market conditions. Overall, the offer bid distortions of production-based CfD affect the DA market supply curve with potential consequences for wind generation and equilibrium prices. In the ID market, both supply and demand curves are affected by offer and bid price distortions. In the following section, we develop a simulation model to study the effect of the derived distortions on aggregate market outcomes based on a case study for the Iberian electricity market. 

\begin{table}[h!]
    \centering
    \footnotesize
    \caption{Summary of bidding strategies}
    \setlength{\extrarowheight}{10pt}
    \setlength{\tabcolsep}{5pt}
\begin{tabular}{@{}lcccc@{}}
        \toprule
         & Quantity bid & \multicolumn{3}{c}{Price bid} \\
         \hline
         & All & \makecell{No CfD} & \makecell{Hourly CfD} & \makecell{Constant CfD}  \\
     \midrule
        \makecell[l]{Day-ahead offer: \\ \textbf{available generation}} & $\hat Q_t^{DA}$ & $c$ &  $p^{DA}_{min}$ & $c-S+\hat v_{\tau}$  \\
  
         \makecell[l]{Intraday offer if $\hat Q_t^{ID}\geq q_t^{DA}$: \\ \textbf{remaining generation}} &  $\hat Q_t^{ID}-q_t^{DA}$ & c  & $c-S+p_t^{DA}$ & $c-S+\hat v_{\tau}$  \\

         \makecell[l]{Intraday bid if $\hat Q_t^{ID}<q_t^{DA}$: \\ \textbf{missing generation}} &  $\hat Q_t^{ID}-q_t^{DA}$ & $\hat p_t^{imb}$   & $\hat p_t^{imb}$ & $\hat p_t^{imb}$ \\

        \makecell[l]{Intraday bid: \\ \textbf{buy-back strategy}} &  $\max\{-q_t^{DA},-\hat Q_t^{ID}\}$ & $c$   & $c-S+p_t^{DA}$ & $c-S+\hat v_{\tau}$ \\
         \bottomrule
    \end{tabular}
    \label{tab:theory_summary}
\end{table}

\section{Methodology}
\label{sec:method}

We quantify the market distortions induced by production-based CfDs using an agent-based sequential market simulator calibrated to historical MIBEL bids for 2025. The simulator reconstructs a DA auction followed by an ID auction, allowing us to disentangle how CfD-induced bidding strategies affect each market sequence.

\subsection{Sequential Market Simulator and Data}

The simulator models the MIBEL market, covering the Spanish and Portuguese bidding zones. It is implemented in Python and reconstructs hourly uniform-price DA and ID auctions. The market-clearing engine uses Gurobi to determine the clearing price $p_t^s$ and accepted volumes $q_t^s$ for auction $s\in\{\text{DA},\text{ID}\}$ by maximising total social welfare from aggregate cumulative supply and demand step functions.

The simulation is initialised with historical price-quantity bids submitted to MIBEL and published by its nominated electricity market operator \citep{omie2025}. Since the raw data include unit identifiers, bid directions, and price-quantity pairs but no explicit technology labels, we classify market units into mutually exclusive technology categories by sequentially querying official databases from Iberian market operators. The detailed classification algorithm is reported in~\ref{anex:class}. We validate the classification by comparing historical wind bids with actual wind generation from the ENTSO-E Transparency Platform \citep{entsoe2026}; total wind DA offers amount to 94\% of observed wind generation in 2025.

\subsection{Scenario Implementation and Parameterisation}
\label{sec:casestudy}

Historical submitted hourly quantities are used as the initial DA capacity forecast of strategic wind onshore units, $\hat Q_t^{DA}$. These units submit DA offers according to the bidding strategies derived in Section~\ref{sec:theory}. After DA clearing, they receive an updated ID capacity forecast, $\hat Q_t^{ID}$, and submit ID offer and demand schedules. Forecast updates are modelled using a uniform stochastic deviation of up to $\pm5\%$ from the cleared DA position \citep{vtt2021}. All non-strategic units retain their historical OMIE bids.

We simulate eleven runs. Scenario~0 reconstructs the historical OMIE order books without bid manipulation and is used for model validation. Scenario~1 is the baseline without a production-based CfD, where strategic wind units bid at marginal cost, set to $c=0$. Scenarios~2--4 represent hourly production-based CfDs with strike prices of 20, 45, and 70\,\euro{}/MWh. Scenarios~5--7 use monthly wind market values $v_m$ as constant reference prices, and Scenarios~8--10 use the annual wind market value $v_y$, again for the same three strike prices. By applying all strategic bidding rules to the same underlying OMIE order book, differences across scenarios can be attributed to the CfD-induced bidding distortions. The full experimental setup is summarised in Table~\ref{tab:experimental_setup}.

\begin{table}[htbp]
\centering
\caption{Experimental setup and parameterisation}
\label{tab:experimental_setup}
\footnotesize
\setlength{\tabcolsep}{3pt}
\renewcommand{\arraystretch}{1.2}
\begin{tabularx}{\columnwidth}{
>{\raggedright\arraybackslash}p{5cm}
>{\centering\arraybackslash}p{2cm}
X}
\toprule
\textbf{Parameter} & \textbf{Symbol} & \textbf{Description} \\
\midrule

Simulation horizon
& --
& Jan.~1--Dec.~31, 2025 (OMIE 2025 dataset, 8,760 hours) \\

Wind strike prices
& $S$
& Low: 20, base: 45, high: 70\,\euro{}/MWh \\

Monthly reference price
& $v_m$
& Monthly wind generation-weighted average DA price (81.18, 102.23, 49.25, 23.42, 21.83, 72.62, 66.42, 68.41, 59.12, 63.21, 52.11 and 71.96\,\euro{}/MWh) \\

Annual reference price
& $v_y$
& Wind generation-weighted average DA price over the full simulation horizon (60.67\,\euro{}/MWh) \\

DA-ID wind forecast deviation
& $\epsilon$
& $\pm 5\%$ (uniform stochastic distribution) \\

Expected imbalance penalty
& $p_t^{imb}\,\forall\, t\in T$
& 500\,\euro{}/MWh (risk-averse threshold enforcing sequential hedging) \\

DA minimum price
& $p_{min}^{DA}$
& $-500$\,\euro{}/MWh (minimum price in MIBEL DA market \citep{omie2025}) \\

\bottomrule
\end{tabularx}
\end{table}

The base strike price is set to 45\,\euro{}/MWh, reflecting a representative levelised cost of electricity (LCOE) \citep{arup2025}. The low and high strike prices, 20 and 70\,\euro{}/MWh, capture sensitivity to technology costs and CfD auction competitiveness. The low value reflects highly competitive recent auctions in Spain and Portugal \citep{schneider2021,iea_019}, while the high value represents undersubscribed German wind auctions in 2023 \citep{bnetza2023}.

In the constant production-based CfD scenarios, the reference price is a technology-specific market value rather than the hourly clearing price. We calculate $v_{\tau}$ as the ex-post generation-weighted average DA market price of the strategic wind portfolio in the baseline scenario:
\begin{align}
\label{eq:ref_value_monthly}
v_{\tau} = \frac{\sum_{t \in \tau} \sum_{u \in U} q_{t,u}^{DA} p_t^{DA}}{\sum_{t \in \tau} \sum_{u \in U} q_{t,u}^{DA}},
\end{align}
where $U$ is the set of strategic wind units, $q_{t,u}^{DA}$ is the cleared DA volume of unit $u$ in hour $t$, and $p_t^{DA}$ is the DA market clearing price. We compute this value over each month, $v_m$, and over the full simulation year, $v_y$. This follows the logic of European market-value-based support schemes, such as the German \textit{Marktwert} design \citep{eeg2023}. The use of baseline values implies that wind generators form expectations based on an undistorted market; limitations of this assumption are discussed below.

An elevated imbalance penalty of 500\,\euro{}/MWh is implemented within the optimisation constraints. Rather than attempting to forecast average realised imbalance costs, this penalty is treated as a behavioural constraint, i.e., an algorithmic ``Big M'' parameter. It prevents the model from substituting physical correction with inexpensive settlement of deviations. Because the objective of the simulation is to isolate cross-market DA-to-ID arbitrage incentives, this conservative threshold enforces risk-averse sequential hedging and confines strategic volume to the simulated ID order book rather than to unmodelled balancing markets.

\subsection{Ex-Post Evaluation Metrics and Settlement}
\label{sec:experimental_design}

As a validation step, we compare the baseline case (Scenario~1) with the historical market reconstruction (Scenario~0). Table~\ref{tab:Res_summary} in~\ref{annex:results} shows that Scenario~1 reproduces Scenario~0 reasonably well, with comparable average DA and ID prices and traded volumes, thereby supporting the validity of the sequential market framework. Small deviations, particularly in curtailment, are expected because observed renewable bids are affected by existing Spanish and Portuguese support schemes, while the baseline scenario applies bids derived under the assumption of purely market-based revenues.

Following the simulated auction sequence, an ex-post realisation phase is conducted to compute three core distortion metrics: price deviation, wind curtailment, and agent profitability. For price deviation, we analyse mean DA and ID prices ($\bar p^{DA}_p$ and $\bar p^{ID}_p$) and their standard deviations ($\sigma^{DA}_p$ and $\sigma^{ID}_p$) as indicators of price volatility. To isolate market distortions induced by CfD arbitrage from exogenous weather forecast errors, we assume that the ID generation forecast is perfectly accurate at delivery, such that actual physical capacity $Q_t$ equals the ID forecast $\hat Q_t^{ID}$.

The final net market position of the agent is given by the sum of its cleared DA and ID volumes, with $q_t^{ID}$ taking a negative value when the agent buys back energy:
\begin{equation}
q_t^{net} = q_t^{DA} + q_t^{ID}.
\end{equation}
Comparing available capacity $Q_t$ with the net market position yields physically delivered energy ($q_t^{del}$), market-based curtailment ($q_t^{curt}$)\footnote{We define market-based curtailment as available output that is not dispatched because the corresponding offer is not cleared in the market or DA commitments are curtailed in favour of ID buy-backs. Curtailment due to grid congestion is not considered in our framework.}, and any remaining imbalance shortfall ($q_t^{short}$):
\begin{align}
q_t^{del} &= \min \{Q_t, q_t^{net}\}, \\
q_t^{curt} &= \max \{0,\, Q_t - q_t^{net}\}, \\
q_t^{short} &= \min \{0,\, Q_t - q_t^{net}\}.
\end{align}
Because we assume perfect ID weather forecasting, any shortfall in the realisation phase is strictly the result of an agent's inability to clear its ID bids.

To disentangle curtailment resulting from the DA and ID markets, we apply a sequential marginal decomposition. First, we calculate the pre-ID curtailment, defined as available energy not cleared in the DA auction:
\begin{equation}
    q_t^{curt,DA} = \max\{0,\, Q_t - q_t^{DA}\}.
\end{equation}
The marginal ID impact is then defined as the difference between final total curtailment and this DA baseline:
\begin{equation}
    q_t^{curt,ID} = q_t^{curt} - q_t^{curt,DA}.
\end{equation}
This definition yields a positive value when agents buy back previously cleared volumes and increase curtailment, and a negative value when the ID market integrates surplus energy and decreases curtailment.

For the final financial settlement, residual shortfalls are valued using a simplified imbalance-price proxy. This choice is motivated by ERSE-reported system service indicators, according to which imbalance-related charges represented approximately 5\% of the spot price in 2025 up to August \citep{apolinario2025}. We therefore set $p_t^{imb}=1.05\,p_t^{DA}$ for the ex-post valuation of uncleared shortfalls. This proxy should not be interpreted as a detailed representation of the imbalance settlement mechanism, which depends on balancing energy activation, imbalance direction, and settlement-period conditions \citep{EU2017Balancing}. By isolating physical market revenues from CfD settlement payments, we calculate the ex-post net profit for each scenario.

The baseline net profit without a production-based CfD is calculated as
\begin{align}
\label{eq:profit_expost_no_cfd}
  \Pi_t =
  q_t^{DA} p_t^{DA}
  + q_t^{ID} p_t^{ID}
  + q_t^{short} p_t^{imb}
  - q_t^{del} c.
\end{align}
The net profit for wind farms subject to a production-based hourly CfD incorporates the hourly reference settlement:
\begin{align}
\label{eq:profit_realised_basic}
    \Pi_t^{hourly} =
    q_t^{DA} p_t^{DA}
    + q_t^{ID} p_t^{ID}
    + q_t^{short} p_t^{imb}
    - q_t^{del} c
    + q_t^{del}(S - p_t^{DA}).
\end{align}
The constant CfD net profit relies on the constant reference market value $v_{\tau}$, where $\tau$ represents either a month or the full simulation horizon:
\begin{align}
\label{eq:profit_realised_2way}
    \Pi_t^{constant} =
    q_t^{DA} p_t^{DA}
    + q_t^{ID} p_t^{ID}
    + q_t^{short} p_t^{imb}
    - q_t^{del} c
    + q_t^{del}(S - v_{\tau}).
\end{align}
By tracking shifts in average DA and ID clearing prices, aggregate market-based curtailment, and profit components across all simulation runs, we quantify and disentangle the structural impact of each CfD design by market sequence.

%------------ FLC short
% \section{Results}
% \label{sec:results}
% This section quantifies how different CfD designs affect market outcomes in the Iberian electricity market under the assumption that all participating wind power plants are subject to the same type of CfD. The analysis focuses on the extent to which strategic curtailment and buy-back incentives induced by distorted offer and bid prices affect results of a day-ahead (DA) followed by an intraday (ID) auction, specifically disentangling the impact of each market sequence. Results are compared between the baseline scenario, where wind power plants are not subject to a production-based CfD and scenarios where wind power plants operate under different types of production-based CfDs. Namely, we consider hourly and constant (monthly and annual) reference prices, each combined with three different strike price levels. First, we present the main mechanisms of distortions at the example of particular hourly auction simulations (Section \ref{sec:hourly}). To assess which of these mechanisms prevails, we also analyse wind generators' average price and quantity bids. Second, we discuss the aggregate effect of the identified distortions on key DA and ID market results (Section \ref{sec:marketoutcomes}). Particularly, we analyse prices, wind curtailment, and ID traded volume as well as implications for generator profits. Finally, we discuss our results and the limitations of our approach (Section \ref{sec:discussion}).
%------------------------

\section{Results}
\label{sec:results}

This section quantifies how production-based CfD designs affect DA and ID market outcomes in the Iberian electricity market. We first illustrate the main distortion mechanisms using selected hourly auctions and average wind bids. We then analyse aggregate effects on prices, wind curtailment, traded volumes, and generator profits, before discussing implications and limitations.

\subsection{Mechanisms of Distortion: Hourly Auctions and Average Bids}
\label{sec:hourly}

This subsection illustrates the main distortion mechanisms using four simulated hourly DA and ID auctions with large ID price deviations from the baseline, identified in Table~\ref{tab:price_diff_identification} in~\ref{annex:results}. We then analyse average wind offer and bid prices and quantities to assess which mechanisms dominate across the full simulation horizon.

\begin{figure}[!h]
    \centering
    \includegraphics[width=0.9\linewidth]{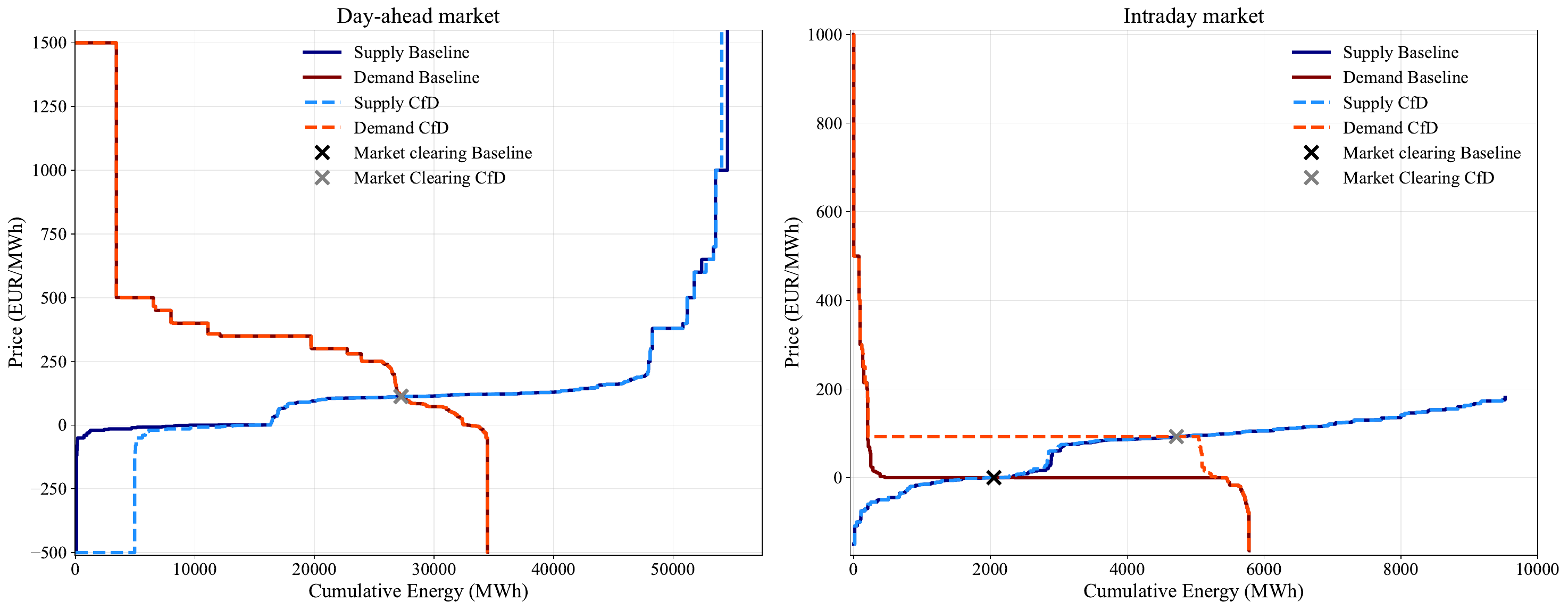}
    \caption{DA and ID supply and demand curves during period 7,218 in the Baseline scenario and the hourly CfD with a strike price of 20 \euro{}/MWh.}
    \label{fig:hourly_cfd_effect}
\end{figure}

\begin{figure}[h!]
    \centering
    \includegraphics[width=0.94\linewidth]{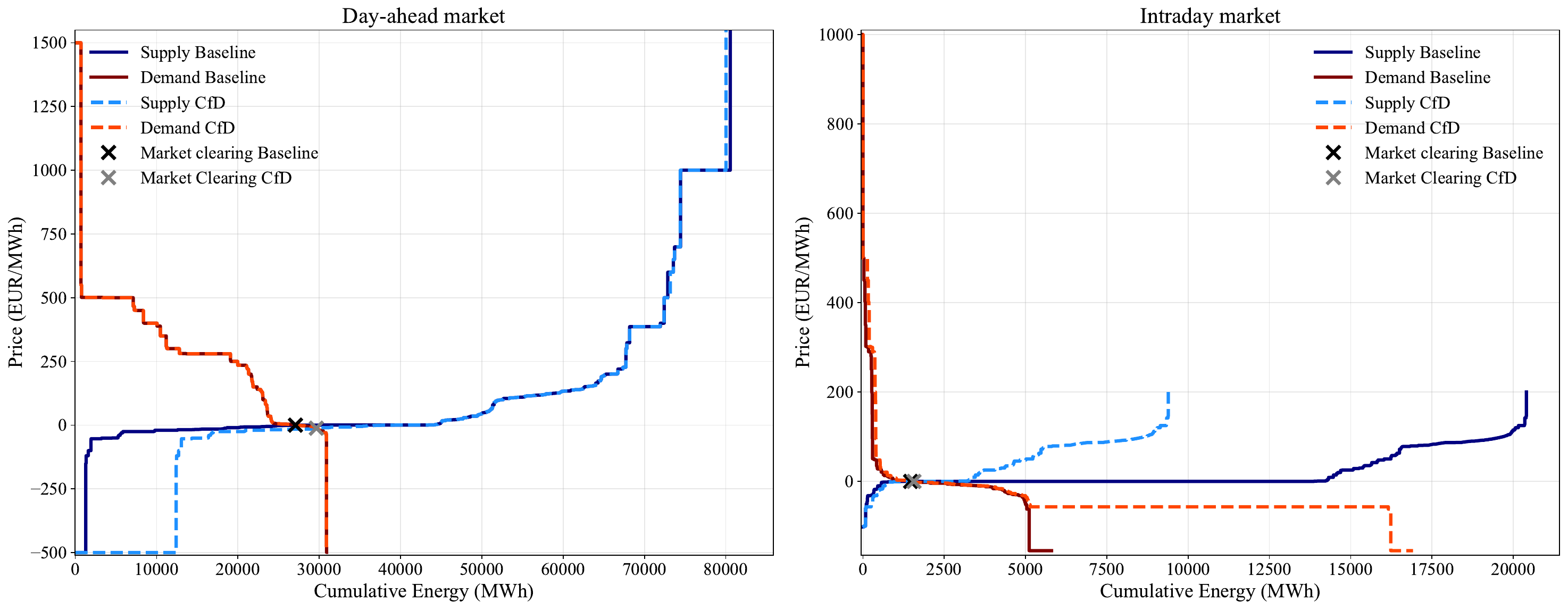}
    \caption{DA and ID supply and demand curves during period 3,327 in the Baseline scenario and the hourly CfD with a strike price of 45 \euro{}/MWh.}
    \label{fig:hourly_quantity_effect}
\end{figure}

\begin{figure}[h!]
    \centering
    \includegraphics[width=0.94\linewidth]{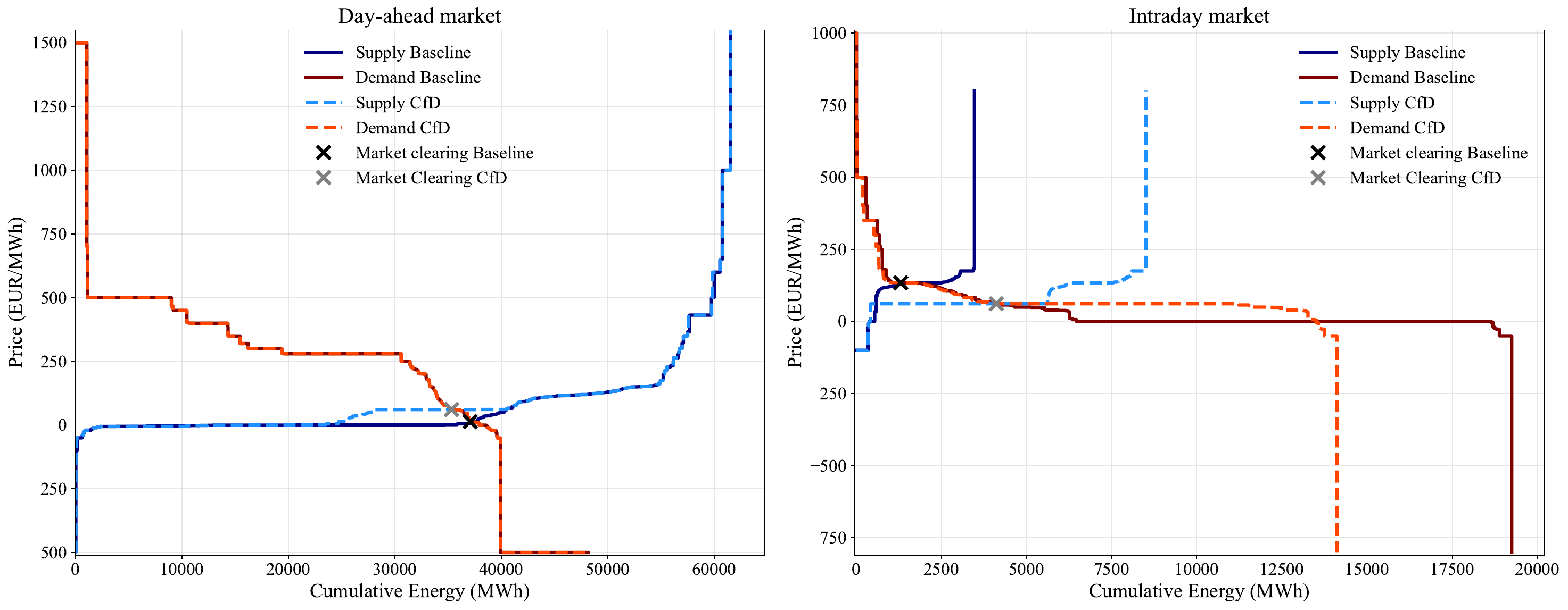}
    \caption{DA and ID supply and demand curves during period 693 in the Baseline scenario and the monthly CfD with a strike price of 20 \euro{}/MWh.}
    \label{fig:monthly_effect}
\end{figure}

\begin{figure}[h!]
    \centering
    \includegraphics[width=0.94\linewidth]{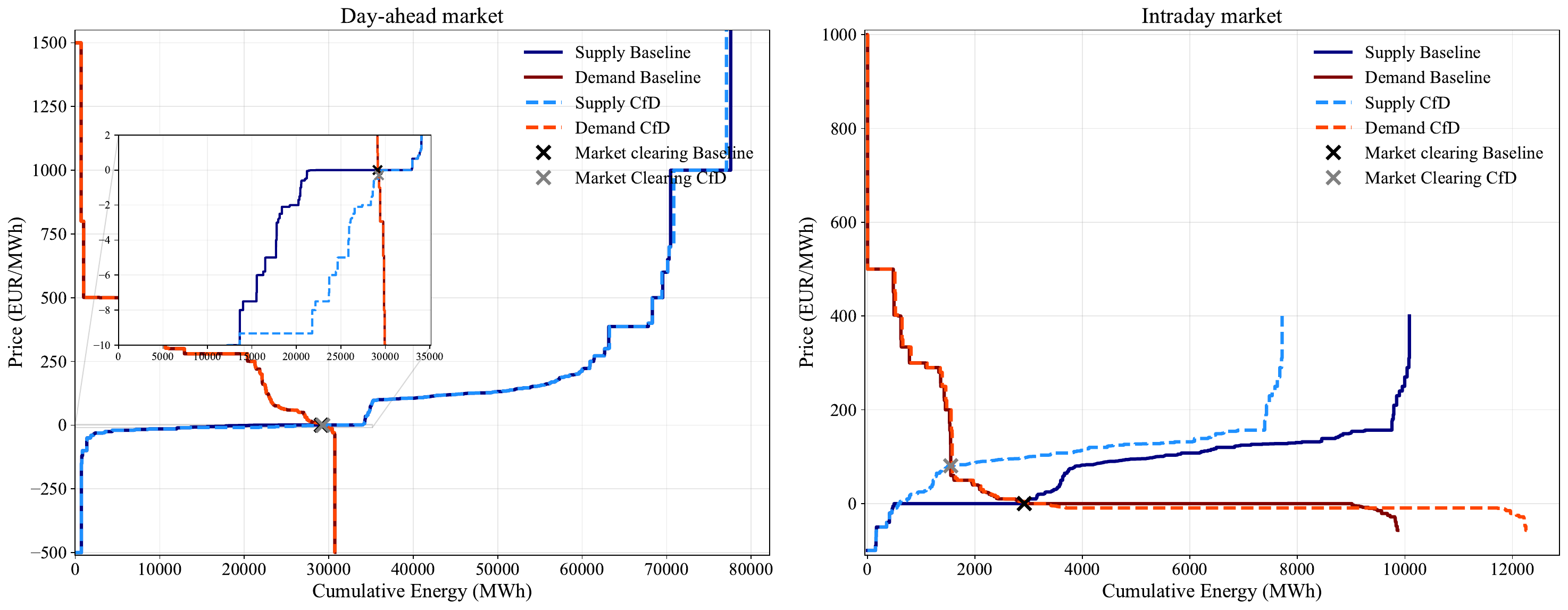}
    \caption{DA and ID supply and demand curves during period 4,785 in the Baseline scenario and the annual CfD with a strike price of 70 \euro{}/MWh.}
    \label{fig:yearly_effect}
\end{figure}

\paragraph{Hourly distortion mechanisms}
Figures~\ref{fig:hourly_cfd_effect}--\ref{fig:yearly_effect} show that production-based CfDs affect ID market outcomes through both vertical and horizontal shifts of wind offer and bid curves. Vertical shifts arise because anticipated CfD payments enter wind generators' ID offers and buy-back bids as virtual marginal costs or revenues. This effect is most visible in Figure~\ref{fig:hourly_cfd_effect}, where a high DA reference price and a low strike price under the hourly CfD make buy-back bids price-setting in the ID market. In this case, wind generators have an incentive to curtail physical generation and repurchase their DA commitments whenever the ID price remains below marginal generation costs plus the expected CfD payment.

Horizontal shifts originate from preceding DA distortions. Figure~\ref{fig:hourly_quantity_effect} shows that lower DA offer prices under the hourly CfD increase DA wind sales, which increases ID demand for missing generation and buy-back volumes. Conversely, Figure~\ref{fig:monthly_effect} shows that higher DA offer prices under the monthly CfD reduce DA wind sales and leave more generation available for ID offers, shifting the ID supply curve to the right. Figure~\ref{fig:yearly_effect} illustrates the opposite case: virtual marginal revenues under the annual CfD increase DA wind sales, reduce remaining ID supply, and create upward pressure on the ID price. Overall, these examples show that ID price distortions are not only caused by altered ID bid prices, but also by DA-induced changes in the quantities submitted to the ID market.

\FloatBarrier

\begin{figure}[!h]
    \centering
    \includegraphics[width=1\linewidth]{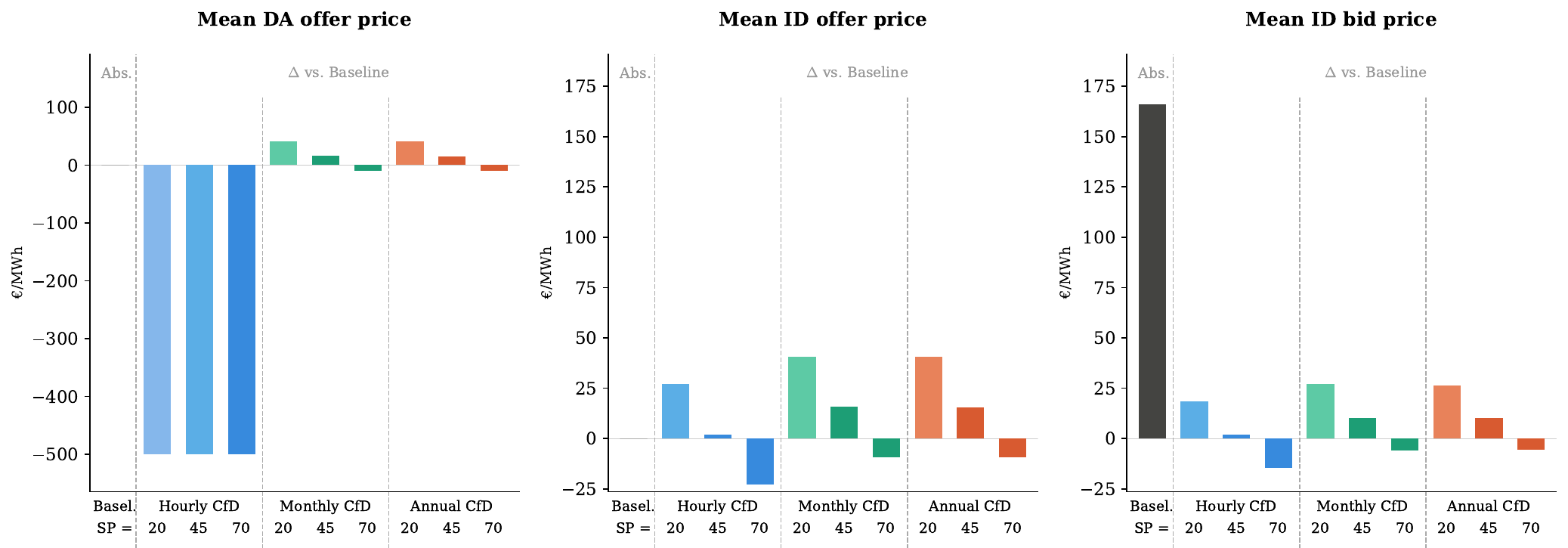}
    \caption{Mean absolute DA and ID offer and bid prices applied in the Baseline (Basel.) scenario and their difference ($\Delta$) to those applied under scenarios with production-based CfD.}
    \label{fig:pricebids}
\end{figure}

\begin{figure}[!h]
    \centering
    \includegraphics[width=1\linewidth]{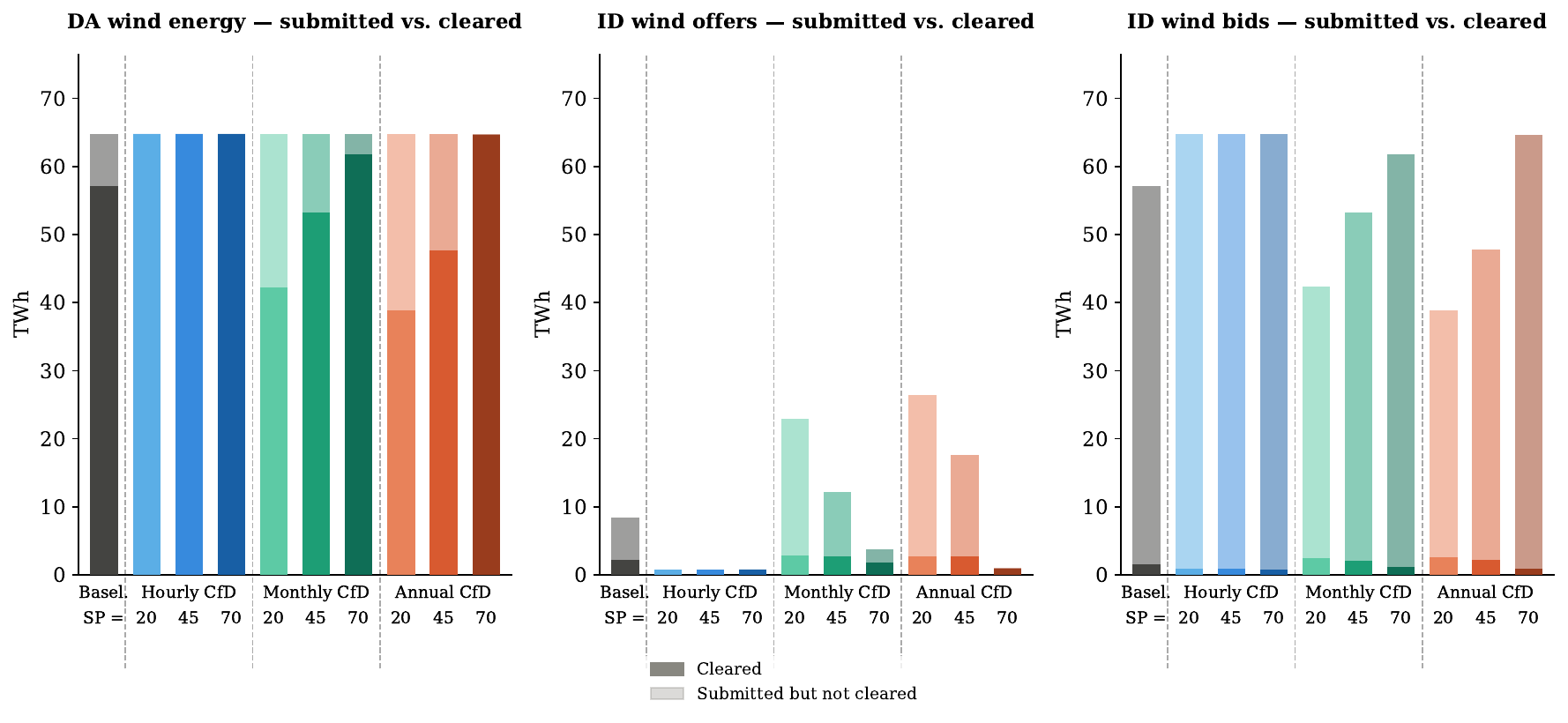}
    \caption{Mean submitted and cleared wind energy offers and bids in the DA and ID markets across the Baseline (Basel.) and nine CfD scenarios.}
    \label{fig:bids}
\end{figure}

\paragraph{Average bids}
Figures~\ref{fig:pricebids} and \ref{fig:bids} assess the prevalence of these mechanisms across all scenarios. Figure~\ref{fig:pricebids} shows that expected CfD payments increase average wind offer and bid prices for strike prices of 20 and 45\,\euro{}/MWh, while expected payments to wind generators lower them for a strike price of 70\,\euro{}/MWh. This pattern applies to DA, ID offer, and ID bid prices under monthly and annual CfDs. Under the hourly CfD, DA offers are always submitted at the minimum price of $-500$\,\euro{}/MWh by construction.

Figure~\ref{fig:bids} shows that DA outcomes determine the quantity shifts observed in the ID market. When CfDs increase DA wind sales, less capacity remains available for ID offers, while ID demand for missing generation and buy-back increases. When CfDs reduce DA wind sales, the opposite occurs: more wind generation remains available for ID offers and ID demand decreases. Across all scenarios, average cleared ID wind offers slightly exceed cleared ID wind bids, indicating that ID wind sales exceed wind purchases on average. Thus, the dominant mechanisms are increased remaining wind generation and increased demand for missing generation, rather than aggregate buy-back-driven curtailment.

\FloatBarrier

\subsection{Aggregate Market Outcomes} 
\label{sec:marketoutcomes}
This subsection analyses how distortions in price and quantity bids affect aggregate market outcomes, disentangling effects resulting from the DA and ID markets.

\begin{figure}[!h]
    \centering
    \includegraphics[width=0.95\linewidth]{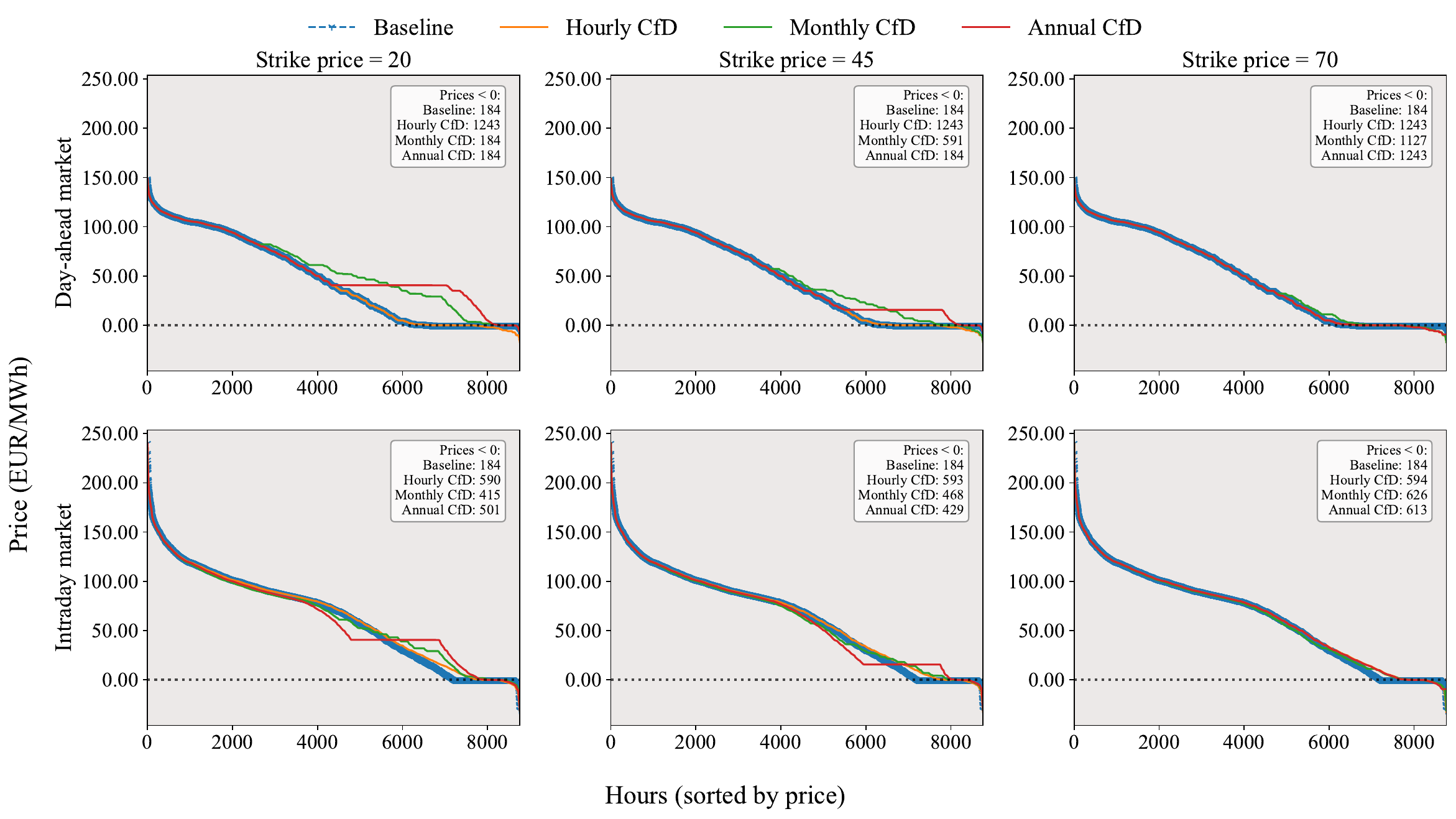}
    \caption{DA and ID price duration curves in the Baseline and nine CfD scenarios by strike price. The inset reports the number of hours with negative prices.}
    \label{fig:price_duration}
\end{figure}

\paragraph{Market clearing prices}
Figure~\ref{fig:price_duration} compares DA and ID price duration curves across CfD scenarios, while Table~\ref{tab:hours_diff} shows the share of hours in which prices are equal to, lower than, or higher than the baseline. Across all CfD designs, ID prices are affected at least as frequently as DA prices, changing in up to 45\% of hours. However, ID price changes occur in both directions and therefore have a smaller effect on annual mean prices. By contrast, DA price effects are more homogeneous within each CfD design, especially when reference prices are held constant. Hourly CfDs mainly affect the lower tail of the DA price distribution, decreasing DA prices in approximately 15\% of hours and increasing the frequency of negative prices, but with only marginal effects on mean DA prices.

\begin{table}[!h]
    \centering
    \caption{Share of hours with DA and ID market prices equal to, lower than, or higher than the Baseline scenario}
    \label{tab:hours_diff}
    \resizebox{\columnwidth}{!}{%
    \small
    \begin{tabular}{ll ccccccccc}
        \toprule
         & Policy & \multicolumn{9}{c}{Production-based CfD} \\
        \midrule
         & Reference Price & \multicolumn{3}{c}{Hourly} & \multicolumn{3}{c}{Monthly} & \multicolumn{3}{c}{Annual} \\
        & Strike Price    & 20 & 45 & 70 & 20 & 45 & 70 & 20 & 45 & 70 \\
        \midrule

     \multirow{3}{*}{\shortstack{\textbf{Day-ahead}\\\textbf{market price}}}
        & $= $ Baseline & 85.76\% & 85.76\% & 85.76\% & 57.97\% & 68.79\% & 79.56\% & 54.76\% & 68.04\% & 85.74\% \\
        & $< $ Baseline & 14.19\% & 14.19\% & 14.19\% & 0.01\% & 6.06\% & 12.62\% & 0.01\% & 0.00\% & 14.19\% \\
        & $> $ Baseline & 0.05\% & 0.05\% & 0.05\% & 42.01\% & 25.15\% & 7.82\% & 45.23\% & 31.96\% & 0.07\% \\
        \midrule
      \multirow{3}{*}{\shortstack{\textbf{Intraday}\\\textbf{market price}}}
        & $= $ Baseline & 79.77\% & 81.58\% & 81.92\% & 57.46\% & 67.76\% & 76.80\% & 55.18\% & 66.13\% & 81.42\% \\
        & $< $ Baseline & 5.18\% & 5.23\% & 5.24\% & 19.29\% & 15.55\% & 11.22\% & 19.96\% & 16.91\% & 5.90\% \\
        & $> $ Baseline & 15.05\% & 13.19\% & 12.84\% & 23.24\% & 16.69\% & 11.98\% & 24.87\% & 16.97\% & 12.67\% \\
        \bottomrule
    \end{tabular}
    }
\end{table}

\begin{figure}[!h]
    \centering
    \includegraphics[width=0.9\linewidth]{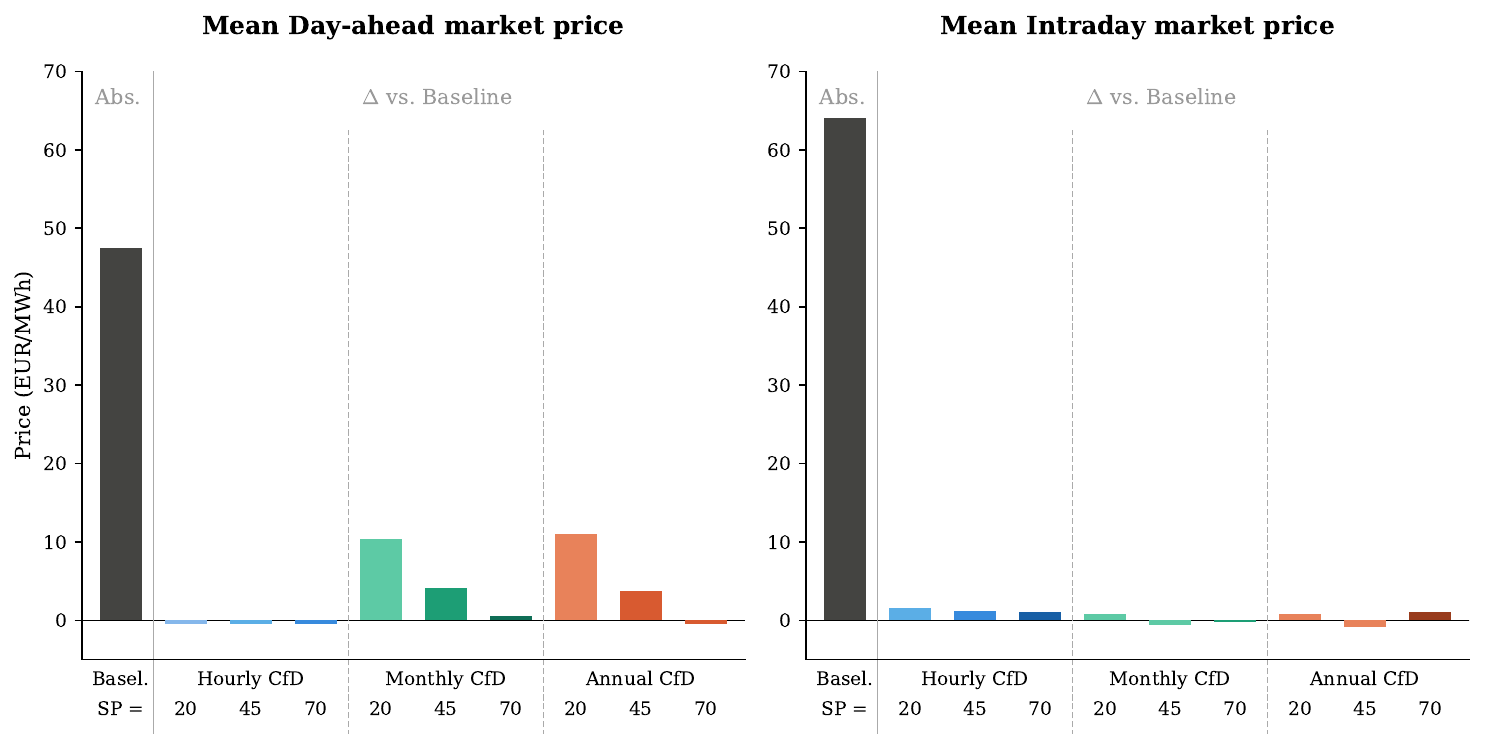}
    \caption{Absolute mean DA and ID market prices in the Baseline scenario and the difference ($\Delta$) of the CfD scenarios' mean prices to the Baseline.}
    \label{fig:prices}
\end{figure}

Figure~\ref{fig:prices} reports deviations in mean DA and ID prices relative to the baseline, with detailed numerical values reported in Table~\ref{tab:Res_summary} in \ref{annex:results}. For constant CfDs, DA and ID price shifts depend on the difference between the reference and strike prices. Low strike prices increase wind offer prices, generating price plateaus and raising DA prices in a large share of hours. This effect is strongest for monthly and annual CfDs with a strike price of 20\,\euro{}/MWh. Higher strike prices reduce or reverse this effect by inducing lower or negative wind offer prices. In the ID market, horizontal shifts of supply and demand curves distribute price effects across both increases and decreases. Consequently, average ID price changes remain limited, even though negative ID prices become more frequent and hourly price increases can reach up to 93\,\euro{}/MWh.

Price volatility follows the same logic. ID price standard deviations decrease across all CfD scenarios, while DA volatility increases in scenarios that induce more negative DA prices, particularly the hourly CfD and the annual CfD with a high strike price.

\FloatBarrier

\begin{figure}[!h]
    \centering
    \includegraphics[width=0.6\linewidth]{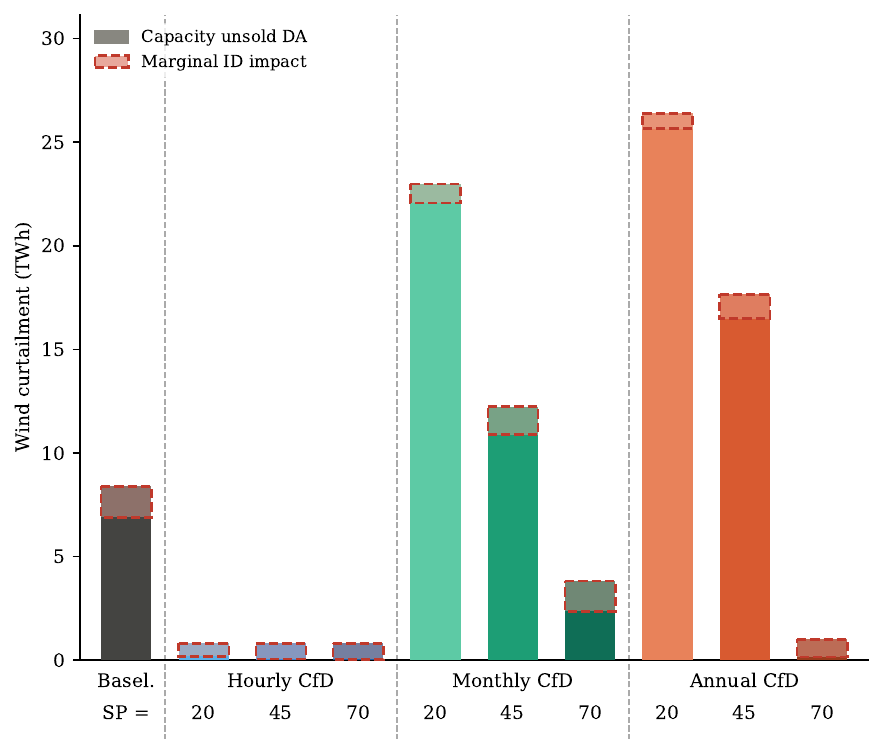}
    \caption{Wind curtailment across the baseline and nine CfD scenarios, decomposed into capacity unsold in the DA market and the marginal ID market impact.}
    \label{fig:idvolume}
\end{figure}

\paragraph{Dispatch and profits}
The effects of production-based CfDs on dispatch are mainly reflected in altered wind curtailment driven by DA market distortions. As shown in Figure~\ref{fig:idvolume}, the hourly CfD and constant CfDs with a high strike price of 70\,\euro{}/MWh almost eliminate market-based wind curtailment by inducing low or negative wind offer prices. Conversely, constant CfDs with low strike prices increase the amount of capacity left unsold in the DA market. The strongest effects occur under the annual and monthly CfDs with a strike price of 20\,\euro{}/MWh, where wind curtailment increases from 6.89~TWh in the baseline to 25.66~TWh and 22.08~TWh, respectively.

Importantly, the decomposition in Figure~\ref{fig:idvolume} shows that curtailment increases are driven by unsold DA capacity, while ID market participation reduces wind curtailment across all CfD scenarios. This indicates that, in aggregate, the ID market partly corrects DA-induced inefficiencies by enabling wind generators to sell remaining available capacity. Under the hourly CfD, this correction almost eliminates curtailment, despite stronger buy-back incentives in some hours. Chronological results for the 45\,\euro{}/MWh strike price, shown in Figure~\ref{fig:area_wind_curt_app} in~\ref{annex:results}, further show that curtailment under constant CfDs follows the temporal pattern of the reference price.

Wind revenue decomposition, shown in Figure~\ref{fig:windprofits_app} in~\ref{annex:results}, indicates that market values are mainly driven by DA revenues, while ID revenues play a minor role. CfD payments claw back revenues above the strike price and compensate revenues below it. As expected, average revenues equal the strike price under the hourly CfD. For monthly and annual CfDs, average revenues deviate from the strike price because payments are decoupled from individual market revenues, indicating heterogeneous wind-unit outcomes that would usually be reflected in different strike prices. Additional results in~\ref{annex:results} show that reduced wind generation is mainly compensated by gas-fired generation, suggesting that higher curtailment may increase emissions. Across technologies, profits mostly follow DA price changes, whereas gas-fired units are more exposed to ID price changes due to their balancing role.

\subsection{Discussion and Limitations}
\label{sec:discussion}

The results are directly relevant for renewable support mechanisms in Spain, Portugal, and other European markets where production-based support schemes linked to realised market outcomes remain common. The analysed two-way CfD structures resemble recent European designs in which payments depend on actual generation volumes and reference market prices.

\paragraph{Policy and market-design Implications}

The simulations show that production-based CfDs distort aggregate outcomes in both DA and ID markets, but through different channels. Wind curtailment and generator profits are mainly determined by DA outcomes, whereas ID prices are affected at least as frequently as DA prices, changing in up to 45\% of annual hours. These ID price effects are particularly relevant for balancing providers, such as gas-fired power plants. These results imply that the operational effects of CfD policies cannot be assessed from DA outcomes alone. As ID and balancing markets become increasingly important for integrating variable renewable generation, CfD design should be evaluated across the full sequence of short-term markets.

The hourly auction analysis and average wind bids show that ID distortions are mainly driven by DA-induced quantity shifts. Changes in DA offer prices alter DA wind sales, which in turn changes the amount of remaining generation offered in the ID market and the volume demanded to cover DA commitments. Although anticipated CfD payments also affect ID bid prices, these price changes mostly occur at the lower tail of the demand curve and become price-setting only in hours with particularly high reference prices. Therefore, buy-back incentives are not strong enough to increase aggregate wind curtailment. Instead, ID market trades reduce curtailment across all CfD scenarios by partially correcting inefficient DA outcomes, although this correction comes at the cost of distorted ID prices.

Importantly, changing the reference-price period within a production-based CfD redistributes rather than eliminates these operational distortions. Hourly reference prices strongly suppress DA price signals and increase the occurrence of negative prices, whereas monthly and annual reference prices can induce substantial wind curtailment and higher DA prices when the reference price exceeds the strike price. The interaction between the strike price and reference price is therefore critical, and longer reference periods should not be regarded as a general solution to the distortions associated with hourly CfDs.

These distortions have implications for consumer costs, generator profits, and emissions. Mean DA prices, used here as an indicator of consumer costs, increase particularly under constant CfDs when wind generators expect to make difference payments to the government. Such situations become more likely when competitive CfD auctions and declining wind LCOE place downward pressure on strike prices. However, payments collected through CfDs could offset higher wholesale prices if redistributed to consumers through a negative levy. In terms of generator profits, inframarginal technologies benefit from higher DA prices, while storage is negatively affected by higher charging costs and lower price volatility. Gas-fired plants are more exposed to ID price changes because of their balancing role. Finally, when higher wind curtailment is compensated by gas-fired generation, emissions are likely to increase.

Overall, our results extend the literature on DA distortions induced by production-based CfDs by providing quantitative evidence of additional ID market effects. While the aggregate magnitude of these ID effects is limited, they may be relevant for providers and buyers of balancing energy. The findings also strengthen the case for production-independent settlement structures, such as capacity- or capability-based CfDs. Although such designs may create different investment incentives, they can reduce operational distortions in sequential electricity markets by removing CfD-induced virtual marginal costs from DA and ID bidding decisions. From a policy-design perspective, where preserving efficient short-term market signals is an objective, such production-independent settlement structures should therefore be considered. Where production-based CfDs are retained, proposed designs should be stress-tested across plausible combinations of strike prices and reference-price periods and across sequential market stages before implementation. Evaluating CfDs only in terms of revenue stabilisation, investment incentives, or DA outcomes may overlook relevant effects on curtailment, negative prices, ID price formation, and balancing providers.

\paragraph{Limitations} 

This work focuses on the distortion mechanisms induced by production-based CfDs in a simplified sequential DA-ID framework. Several real-market complexities are intentionally neglected to isolate the effects under analysis.

First, the study models one ID auction and does not explicitly represent all six MIBEL ID auctions or continuous ID trading. Including additional ID trading opportunities could increase the ability of wind generators to sell remaining capacity after the first ID auction. Second, balancing markets and imbalance settlement mechanisms are represented only through a simplified ex-post proxy. In practice, participants may benefit from imbalance positions opposite to the system imbalance, which could partially offset some incentives identified in this work. Third, the analysis assumes perfectly competitive and uniform bidding behaviour for the strategic wind portfolio, while all other technologies retain their historical bids. This abstraction helps isolate the CfD effect but may amplify simulated DA-ID price differences relative to real markets. Fourth, constant-reference CfD strategies require wind generators to forecast reference prices. We use baseline market values for this purpose, whereas in practice generators may update expectations throughout the year and account for strategic behaviour by other participants. Finally, the analysis abstracts from investment incentives associated with different CfD designs, which could affect plant location, technology choices, market values, and price volatility.
Future work should test the robustness of these findings using multiple weather years, continuously updated reference-price expectations, a more detailed representation of balancing markets and continuous ID trading, and endogenous investment responses to different CfD designs.

\section{Conclusion and Policy Implications}
\label{sec:conclusion}

Contracts for Difference (CfDs) are now the mandatory support scheme for renewables in the European Union \citep{eu2024}. CfDs that settle payments based on actual renewable generation are known to distort dispatch decisions with important implications for the efficiency of electricity markets. This paper provides a novel quantitative assessment of these distortions in a sequential electricity market, disentangling their effects on day-ahead (DA) and intraday (ID) auctions. We derive optimal bidding strategies for wind generators under three settings: no production-based CfD, a production-based CfD with an hourly reference price, and a production-based CfD with a constant monthly or annual reference price. Using historical bid data from the Iberian electricity market in 2025, we simulate strategic wind bidding under these designs and quantify their aggregate market impacts across three strike price levels.

The theoretical analysis shows that production-based CfDs affect renewable bidding through two channels. In the DA market, CfD payments alter the opportunity cost of generation and therefore change offer prices. In the ID market, they affect both buy-back incentives and the willingness to sell generation made available by updated forecasts. These distortions arise whenever accepted trades affect not only generation costs but also anticipated CfD payments. By contrast, bids to purchase missing generation and avoid imbalance payments remain unaffected.

The simulation results show that wind curtailment is primarily driven by DA market effects. Hourly production-based CfDs almost eliminate wind curtailment by removing DA price signals, while constant-reference CfDs reduce curtailment when wind generators expect to receive CfD payments. Conversely, curtailment increases when generators expect to make payments to the CfD counterparty, a situation that becomes more likely under lower strike prices. Across all scenarios, ID market activity partly offsets DA-induced curtailment by enabling wind generators to sell remaining available generation.

ID prices, however, are affected at least as frequently as DA prices, changing in up to 45\% of annual hours. DA price effects are more homogeneous within each CfD design, while ID prices change in both directions due to vertical and horizontal shifts of ID supply and demand curves. Mean DA prices increase by up to 10\,\euro{}/MWh, whereas ID prices show more limited average effects but larger hourly deviations, including more frequent negative prices and increases of up to 93\,\euro{}/MWh. Taken together, these results show that changing the reference-price period within a production-based CfD redistributes rather than eliminates operational distortions: different designs shift the effects across curtailment, negative prices, and DA and ID price formation, with their magnitude depending critically on the relationship between the strike and reference prices.

The identified distortions also affect generator profitability and system outcomes. Inframarginal generators' profits generally follow DA price changes, while gas-fired power plants are more exposed to ID price effects due to their balancing role. Higher wind curtailment is mainly compensated by gas-fired generation, suggesting potential emission increases. CfD payments claw back revenues above the strike price and compensate revenues below it, but average revenues under monthly and annual CfDs deviate from strike prices because settlement is decoupled from individual market revenues.

From a policy perspective, the results support the case for production-independent CfD designs, such as capacity- or capability-based contracts, which decouple settlement from physical generation and avoid introducing virtual marginal costs into short-term bidding. Where preserving efficient short-term market signals is a policy objective, such settlement structures should therefore be considered alongside conventional production-based designs. Where production-based CfDs are retained, policymakers should assess proposed contracts across plausible combinations of strike prices and reference-price periods rather than treating the reference period as an isolated design choice. In addition, their operational effects should be evaluated across sequential DA, ID, and, where relevant, balancing markets, since focusing on DA outcomes alone may overlook substantial downstream distortions. As renewable penetration increases and ID markets become more important for balancing variable generation, these operational distortions should be carefully considered in the design of future renewable support schemes and private hedging instruments such as power purchase agreements. More broadly, CfD evaluation should complement conventional objectives such as investment de-risking and revenue stabilisation with an explicit assessment of short-term market efficiency. Future work should extend the analysis to multiple weather years, continuous ID and balancing markets, and strategic multi-agent bidding behaviour.

%TC:ignore 
\section*{CRediT authorship contribution statement}
\textbf{Silke Johanndeiter}: Conceptualization, Methodology, Formal analysis, Visualization, Writing - Original Draft. 
\textbf{Ricardo Faia}: Conceptualization, Methodology, Software, Data curation, Visualization, Writing - Original Draft. 
\textbf{Fernando Lezama}: Conceptualization, Methodology, Software, Formal analysis, Writing - Original Draft. 
\textbf{Zita Vale}: Validation, Resources, Writing – review \& editing,

%\section*{Data availability statement} 
%The data and code that support the findings of this study are openly available in public repositories at XXXX.
\section*{Data availability statement}
Data will be made available on request.

\section*{Acknowledgements}
%GECAD authors acknowledge the work facilities and equipment provided by GECAD research center (UID/00760/2025), DOI: https://doi.org/10.54499/UID/00760/2025 and (UID/PRR/00760/2025), DOI: https://doi.org/10.54499/UID/PRR/00760/2025 to the project team. Ricardo Faia is supported by FCT with grant 2024.10380.CEECIND and Fernando Lezama by FCT with grant 2024.11606.CEECIND.
The authors acknowledge the work facilities and equipment provided by GECAD research center (UID/00760/2025, DOI \href{https://doi.org/10.54499/UID/00760/2025}{https://doi.org/10.54499/UID/00760/2025}) to the project team. Ricardo Faia is supported by FCT with grant 2024.10380.CEECIND and Fernando Lezama by FCT with grant 2024.11606.CEECIND. Silke Johanndeiter's contribution to this paper has been funded by the project REWARDS, which receives funding from the German Federal Ministry for Economic Affairs and Energy (BMWE) under funding reference numbers 03EI1083A, 03EI1083B, 03EI1083C, 03EI1083D.

\section*{Declaration of generative AI and AI-assisted technologies in the manuscript preparation process}   
During the preparation of this work, the authors used ChatGPT and Claude to support data visualisation code and language proofing. The authors reviewed and edited the output as needed and take full responsibility for the content of the published article. 

\newpage

\appendix
\section{Theory: Additional material}
\subsection{Numerical examples}
\label{app:numerical}

The underlying logic of the \textit{buy-back} strategy can be illustrated using a simple numerical example. Consider a power plant that sells $q_t^{DA}=1\,\text{MWh}$ in the DA market at $p_t^{DA}=50\,\euro{}/\text{MWh}$ and has marginal generation costs of $c=1\,\euro{}/\text{MWh}$. If the plant does not adjust its position in the ID market and physically generates the committed energy, its profit without a CfD is
\begin{align}
1\,\text{MWh}\cdot(50-1)\,\euro{}/\text{MWh}=49\,\euro{}.
\end{align}
Suppose instead that the plant buys back $1\,\text{MWh}$ in the ID market at $p_t^{ID}=0.9\,\euro{}/\text{MWh}$ and curtails physical output. It still receives the DA revenue but avoids generation costs, so its profit becomes
\begin{align}
1\,\text{MWh}\cdot(50-0.9)\,\euro{}/\text{MWh}=49.1\,\euro{}.
\end{align}
The plant therefore benefits from buying back its DA commitment whenever $p_t^{ID}<c$.

The example can be extended to a production-based hourly CfD. Consider the same DA sale, DA price and marginal cost, and assume a strike price of $S=30\,\euro{}/\text{MWh}$. Since the hourly reference price is $p_t^{DA}=50\,\euro{}/\text{MWh}$, the plant pays $p_t^{DA}-S=20\,\euro{}/\text{MWh}$ to the CfD counterparty for each MWh physically generated. If the plant generates the DA commitment and does not trade in the ID market, its profit is
\begin{align}
1\,\text{MWh}\cdot(50-1+30-50)\,\euro{}/\text{MWh}=29\,\euro{}.
\end{align}

Now assume that the updated ID forecast is $\hat Q_t^{ID}=1.1\,\text{MWh}$, so the plant has an additional $0.1\,\text{MWh}$ available for ID sales. Selling this additional energy is profitable only if the ID price covers both marginal generation costs and the additional CfD payment obligation. The marginal profit from selling $0.1\,\text{MWh}$ in the ID market is
\begin{align}
0.1\,\text{MWh}\cdot(p_t^{ID}-c+S-p_t^{DA}).
\end{align}
For example, if $p_t^{ID}=21.1\,\euro{}/\text{MWh}$, the additional profit is
\begin{align}
0.1\,\text{MWh}\cdot(21.1-1+30-50)\,\euro{}/\text{MWh}
=
0.01\,\euro{}.
\end{align}
Hence, additional ID sales are profitable when $p_t^{ID}>c-S+p_t^{DA}=21\,\euro{}/\text{MWh}$.

The same threshold determines the buy-back strategy. If the plant buys back $1\,\text{MWh}$ in the ID market and curtails physical output, it keeps the DA revenue but avoids both generation costs and the CfD payment. For an ID price of $p_t^{ID}=20.9\,\euro{}/\text{MWh}$, its profit becomes
\begin{align}
1\,\text{MWh}\cdot(50-20.9)\,\euro{}/\text{MWh}=29.1\,\euro{}.
\end{align}
This is higher than the $29\,\euro{}$ obtained by physically generating the DA commitment. Therefore, the plant benefits from buying back its DA commitment whenever $p_t^{ID}<c-S+p_t^{DA}=21\,\euro{}/\text{MWh}$.

If the strike price is higher than the reference price, the direction of the CfD payment changes. For example, with $S=70\,\euro{}/\text{MWh}$ and $p_t^{DA}=50\,\euro{}/\text{MWh}$, the plant receives $20\,\euro{}/\text{MWh}$ for each MWh generated. In that case, buying back the DA commitment is profitable only if the ID price is below
\begin{align}
c-S+p_t^{DA}=1-70+50=-19\,\euro{}/\text{MWh}.
\end{align}
Thus, the ID price must be sufficiently negative to compensate the plant for the foregone positive CfD payment.

For a production-based CfD with a constant reference price, the same logic applies after replacing the hourly DA reference price $p_t^{DA}$ by the expected constant reference price $\hat v_{\tau}$. The buy-back threshold is therefore $p_t^{ID}<c-S+\hat v_{\tau}$. The same numerical values as in the hourly example are obtained only if $\hat v_{\tau}=50\,\euro{}/\text{MWh}$; if instead $\hat v_{\tau}=30\,\euro{}/\text{MWh}$ and $S=30\,\euro{}/\text{MWh}$, the CfD payment is zero and the buy-back threshold reduces to $p_t^{ID}<c$.

\subsection{Derivations of price thresholds for bidding strategies}
\label{sec:derivation}

\subsubsection{No production-based CfD}

\paragraph{Case 1: ID forecast is higher than DA sales}
If $\hat Q_t^{ID}\geq q_t^{DA}$, the imbalance term is zero for additional ID sales. For an ID offer $(q_n,p_n)^{S,ID}$, expected profits are
\begin{align}
 \hat \Pi_t^{ID,\text{none}}
 =
 q_n p_n
 +
 q_t^{DA}p_t^{DA}
 -
 (q_t^{DA}+q_n)c .
\end{align}
Therefore,
\begin{align}
\frac{\partial \hat \Pi_t^{ID,\text{none}}}{\partial q_n}=p_n-c,
\end{align}
so selling additional generation is profitable when $p_n\geq c$. The optimal ID offer is therefore $(q_1,p_1)^{S,ID}=(\hat Q_t^{ID}-q_t^{DA},c)$.

For an ID buy-back bid, profits increase when the cost of buying energy ID is lower than the avoided generation cost:
\begin{align}
-q_n p_n + q_n c > 0
\quad \Leftrightarrow \quad
p_n < c .
\end{align}
Thus, the buy-back bid is $(q_1,p_1)^{D,ID}=(-q_t^{DA},c)$.

\paragraph{Case 2: ID forecast is lower than DA sales}
If $\hat Q_t^{ID}<q_t^{DA}$, the plant can first buy the missing generation, $q_1=\hat Q_t^{ID}-q_t^{DA}<0$, to avoid imbalance payments. This bid is profitable when
\begin{align}
&(\hat Q_t^{ID}-q_t^{DA})p_n
+
q_t^{DA}p_t^{DA}
-
\hat Q_t^{ID}c
\nonumber\\
&>
q_t^{DA}p_t^{DA}
-
\hat Q_t^{ID}c
+
(\hat Q_t^{ID}-q_t^{DA})\hat p_t^{imb}
\nonumber\\
&\Leftrightarrow
p_n < \hat p_t^{imb}.
\end{align}
Hence, the missing-generation bid is $(q_1,p_1)^{D,ID}=(\hat Q_t^{ID}-q_t^{DA},\hat p_t^{imb})$.

After the missing generation has been bought, the plant may also buy back the remaining DA commitment. The remaining buy-back quantity is $q_2=-\hat Q_t^{ID}$. This is profitable when
\begin{align}
-\hat Q_t^{ID}p_2+\hat Q_t^{ID}c>0
\quad \Leftrightarrow \quad
p_2<c.
\end{align}
Thus, the second ID demand bid is $(q_2,p_2)^{D,ID}=(-\hat Q_t^{ID},c)$.

\subsubsection{Production-based CfD with hourly reference price}

Under the hourly production-based CfD, the same missing-generation bid applies because buying missing generation avoids imbalance payments but does not change the CfD settlement margin. For $q_1=\hat Q_t^{ID}-q_t^{DA}<0$:
\begin{align}
&(\hat Q_t^{ID}-q_t^{DA})p_n
+
q_t^{DA}p_t^{DA}
-
\hat Q_t^{ID}(c-S+p_t^{DA})
\nonumber\\
&>
q_t^{DA}p_t^{DA}
-
\hat Q_t^{ID}(c-S+p_t^{DA})
+
(\hat Q_t^{ID}-q_t^{DA})\hat p_t^{imb}
\nonumber\\
&\Leftrightarrow
p_n < \hat p_t^{imb}.
\end{align}
Hence, the missing-generation bid remains $(q_1,p_1)^{D,ID}=(\hat Q_t^{ID}-q_t^{DA},\hat p_t^{imb})$.

The buy-back bid, however, is affected by CfD payments because buying energy ID allows the plant to avoid physical generation and therefore avoid or forego CfD payments. For the remaining buy-back quantity $q_2=-\hat Q_t^{ID}$, buying back is profitable when
\begin{align}
-\hat Q_t^{ID}p_2+\hat Q_t^{ID}(c-S+p_t^{DA})>0
\quad \Leftrightarrow \quad
p_2<c-S+p_t^{DA}.
\end{align}
Thus, the hourly CfD buy-back bid is $(q_2,p_2)^{D,ID}=(-\hat Q_t^{ID},c-S+p_t^{DA})$.

%--------------------

\subsubsection{Production-based CfD with constant reference price}

The derivation for the constant production-based CfD is identical to the hourly case after replacing $p_t^{DA}$ by $\hat v_{\tau}$. Hence, the buy-back threshold becomes $p_2<c-S+\hat v_{\tau}$, while the missing-generation bid remains priced at $\hat p_t^{imb}$.

\section{Method: Detailed unit classification algorithm}
\label{anex:class}
The algorithm resolves unit classifications by sequentially querying the official databases of the Iberian market and system operators:
\begin{enumerate}
    \item OMIE (Operador del Mercado Ibérico de Energía) unit and agents registration lists\footnote{https://www.omie.es/pt/lista-de-agentes}.
    \item REE (Red Eléctrica de España) programming unit registries\footnote{http://www.esios.ree.es/es/unidades-fisicas}.
    \item REN (Redes Energéticas Nacionais) Portuguese market agents list\footnote{https://mercado.ren.pt/PT/Electr/InfoMercado/InfStructMerc/Paginas/AgentesMerc.aspx}.
\end{enumerate}
For each unit in the simulation dataset, the algorithm executes the following sequence:
\begin{itemize}
    \item OMIE Registry Mapping: If the unit code is present in the OMIE unit database, classification is determined using a keyword-matching pipeline over the official OMIE technology (\texttt{TECNOLOGIA}), unit type (\texttt{TIPO\_UNIDAD}), and description (\texttt{DESCRIPCION}) fields.
    \item REE Registry Mapping: If the unit is not resolved via OMIE but is registered in Spain's REE database, it is classified using REE's production type (\texttt{TIPO\_PRODUCCION}), unit type (\texttt{TIPO\_UP}), and short/long descriptions.
    \item REN Registry Mapping: For Portuguese units not registered in the Spanish databases, the algorithm queries the REN database, separating assets by regime---Special Regime (primarily renewable/cogeneration) and Ordinary Regime (large hydro and thermal units), using their operational type (\texttt{Tipo}) and description.
    \item Fallback and Imputation: Units not matching any operator database default to their legacy classifications. Bids from consumer units are mapped to \texttt{LOAD}, while offers from producer units are mapped to ``Unknown Sellers''.
\end{itemize}
The classification categories are mapped using specific textual heuristics:
\begin{itemize}
    \item Solar Differentiation: Solar units are classified as \textit{Solar CSP} if their database record contains keywords like \texttt{SOLAR TÉRMICA}, \texttt{CSP}, or \texttt{TERMOSOLAR}. Other solar units are classified as \textit{Solar PV large}, unless the description contains residential keywords (\texttt{RESIDENCIAL} or \texttt{AUTOCONSUMO}), in which case they are classified as \textit{Solar PV residential}.
    \item Hydro and Pumped Storage (PHS) Separation: Hydraulic units containing pumping keywords (\texttt{BOMBEO} or \texttt{BOMBAGEM}) are mapped to storage categories. If the transaction type is a buy bid, they are classified as \textit{PHS Charge} (or \textit{Batteries Charge} if battery keywords are found), otherwise they are classified as \textit{PHS Discharge} (or \textit{Batteries Discharge}). Standard hydraulic units containing flow keywords (\texttt{PASADA} or \texttt{FLUYENTE}) are classified as \textit{Run-of-River (ROR)}, while standard reservoirs are classified as \textit{Hydro Discharge}.
    \item Combined Cycle Gas: Mapped to \textit{Gas} based on keywords such as \texttt{CICLO COMBINADO}, \texttt{GAS NATURAL}, or specific plant designations.
    \item Thermal Renewables \& Non-Renewables: Renewables containing \texttt{biomass} or \texttt{biogas} keywords map to \textit{Biomass} and \textit{Biofuel}, while others default to \textit{Others renewable}. Non-renewable thermal units (coal, fuel oil, waste) map to \textit{Others non-renewable} and units whose records contain the keyword \texttt{Nuclear} are classified as \textit{Nuclear}.
\end{itemize}
%We verify our unit classification by comparing total results of the DA market simulation for Scenario 0 (historical bids) to actual generation (cf. Figure \ref{fig:wind_comparision}). Identified wind quantity bids submitted to the DA market amount 94\% of actual wind generation. match actual generation 
%\begin{figure}[!ht]
 %   \centering
   % \includegraphics[width=0.8\linewidth]{Figures/Wind_Onshore_Comparison_Monthly.pdf}
  %  \caption{Total Wind Generation resulting from the Day-ahead market (DAM) simulation of Scenario 0 (historical bidding data) compared to wind generation data from ENTSO-E}
  %  \label{fig:wind_comparision}
%\end{figure}
\newpage
\section{Additional Results}
\label{annex:results}
\begin{table*}[!ht]
\centering
\caption{Most extreme intraday price deviations from Baseline by scenario.}
\label{tab:price_diff_identification}
\small
\setlength{\tabcolsep}{3pt}
\begin{tabular}{l cccc cccc}
\toprule
& \multicolumn{4}{c}{\textbf{Day-ahead market}} & \multicolumn{4}{c}{\textbf{Intraday market}} \\
\cmidrule(lr){2-5} \cmidrule(lr){6-9}
\textbf{Scenario} & Period & $p_{\text{Baseline}}$ & $p$ & $\Delta p$ & Period & $p_{\text{Baseline}}$ & $p$ & $\Delta p$ \\
& & (\euro{}/MWh) & (\euro{}/MWh) & (\euro{}/MWh) & & (\euro{}/MWh) & (\euro{}/MWh) & (\euro{}/MWh) \\
\midrule
\multicolumn{9}{l}{\textit{Hourly CfD}} \\
\quad SP = 20 & 3329 & $-$0.01 & $-$15.00 & $-$14.99 & 7218 &   0.00 &  92.64 &  92.64 \\
\quad SP = 45 & 3329 & $-$0.01 & $-$15.00 & $-$14.99 & 4785 &   0.00 &  80.54 &  80.54 \\
\quad SP = 70 & 3329 & $-$0.01 & $-$15.00 & $-$14.99 & 4785 &   0.00 &  80.54 &  80.54 \\
\midrule
\multicolumn{9}{l}{\textit{Monthly CfD}} \\
\quad SP = 20 & 1336 &   0.00 &  82.23 &  82.23 &  693 & 133.85 &  61.18 & $-$72.67 \\
\quad SP = 45 & 1237 &   0.00 &  57.23 &  57.23 &  717 & 115.00 &  36.18 & $-$78.82 \\
\quad SP = 70 & 1096 &   0.00 &  32.23 &  32.23 &  717 & 115.00 &  11.18 & $-$103.82 \\
\midrule
\multicolumn{9}{l}{\textit{Annual CfD}} \\
\quad SP = 20 &  122 &   0.00 &  40.67 &  40.67 & 2060 & 148.00 &  40.67 & $-$107.33 \\
\quad SP = 45 &  122 &   0.00 &  15.67 &  15.67 &  717 & 115.00 &  15.67 &  $-$99.33 \\
\quad SP = 70 & 2602 &   0.00 & $-$9.33 & $-$9.33 & 4785 &   0.00 &  80.54 &  80.54 \\
\bottomrule
\end{tabular}
\end{table*}

\begin{table*}[!ht]
\centering
\caption{Overview of key numerical results by scenario}
\resizebox{\columnwidth}{!}{%
\label{tab:Res_summary}
\begin{tabular}{lcccccccccccc}
\toprule
\textbf{Scenario} & & \textbf{0} & \textbf{1} & \textbf{2} & \textbf{3} & \textbf{4} & \textbf{5} & \textbf{6} & \textbf{7} & \textbf{8} & \textbf{9} & \textbf{10} \\
\midrule
Policy & & \multicolumn{2}{c}{Historical Baseline} & \multicolumn{9}{c}{Production-based CfD} \\
Strike price & & -- & -- & 20 & 45 & 70 & 20 & 45 & 70 & 20 & 45 & 70 \\
Reference price & & -- & -- & \multicolumn{3}{c}{Hourly} & \multicolumn{3}{c}{Monthly} & \multicolumn{3}{c}{Annual} \\
\midrule
$\bar{p}^{DA}$ & (\euro{}/MWh)  & 48.30 & 47.49 & 47.04 & 47.04 & 47.04 & 57.93 & 51.62 & 48.02 & 58.47 & 51.22 & 47.06 \\
$\sigma_p^{DA}$ & (\euro{}/MWh) & 43.04 & 43.31 & 43.83 & 43.83 & 43.83 & 37.17 & 40.42 & 43.04 & 34.43 & 39.60 & 43.80 \\ 
$\bar{p}^{ID}$ & (\euro{}/MWh) & 64.30 & 64.04 & 65.57 & 65.23 & 65.10 & 64.85 & 63.50 & 63.81 & 64.82 & 63.21 & 65.06 \\ 
$\sigma_p^{ID}$ & (\euro{}/MWh) & 47.44 & 47.67 & 46.58 & 46.64 & 46.72 & 44.53 & 46.19 & 47.07 & 44.31 & 46.91 & 46.73 \\
ID Volume & (TWh) & 33.74 & 29.23 & 27.20 & 27.07 & 27.05 & 31.69 & 30.80 & 28.74 & 31.54 & 31.06 & 27.16 \\ 
Wind curtailment & (TWh) & 6.29 & 6.89 & 0.18 & 0.05 & 0.03 & 22.08 & 10.89 & 2.35 & 25.66 & 16.48 & 0.15 \\ 
Capacity unsold DA & (TWh) & 6.19 & 8.38 & 0.81 & 0.81 & 0.81 & 22.99 & 12.24 & 3.82 & 26.39 & 17.65 & 1.00 \\
Marginal ID impact & (TWh) & 0.10 & -1.49 & -0.63 & -0.75 & -0.78 & -0.92 & -1.36 & -1.47 & -0.73 & -1.17 & -0.85 \\ 
\bottomrule
\end{tabular}
}
\end{table*}

\begin{figure}[!ht]
    \centering
    \includegraphics[width=1\linewidth]{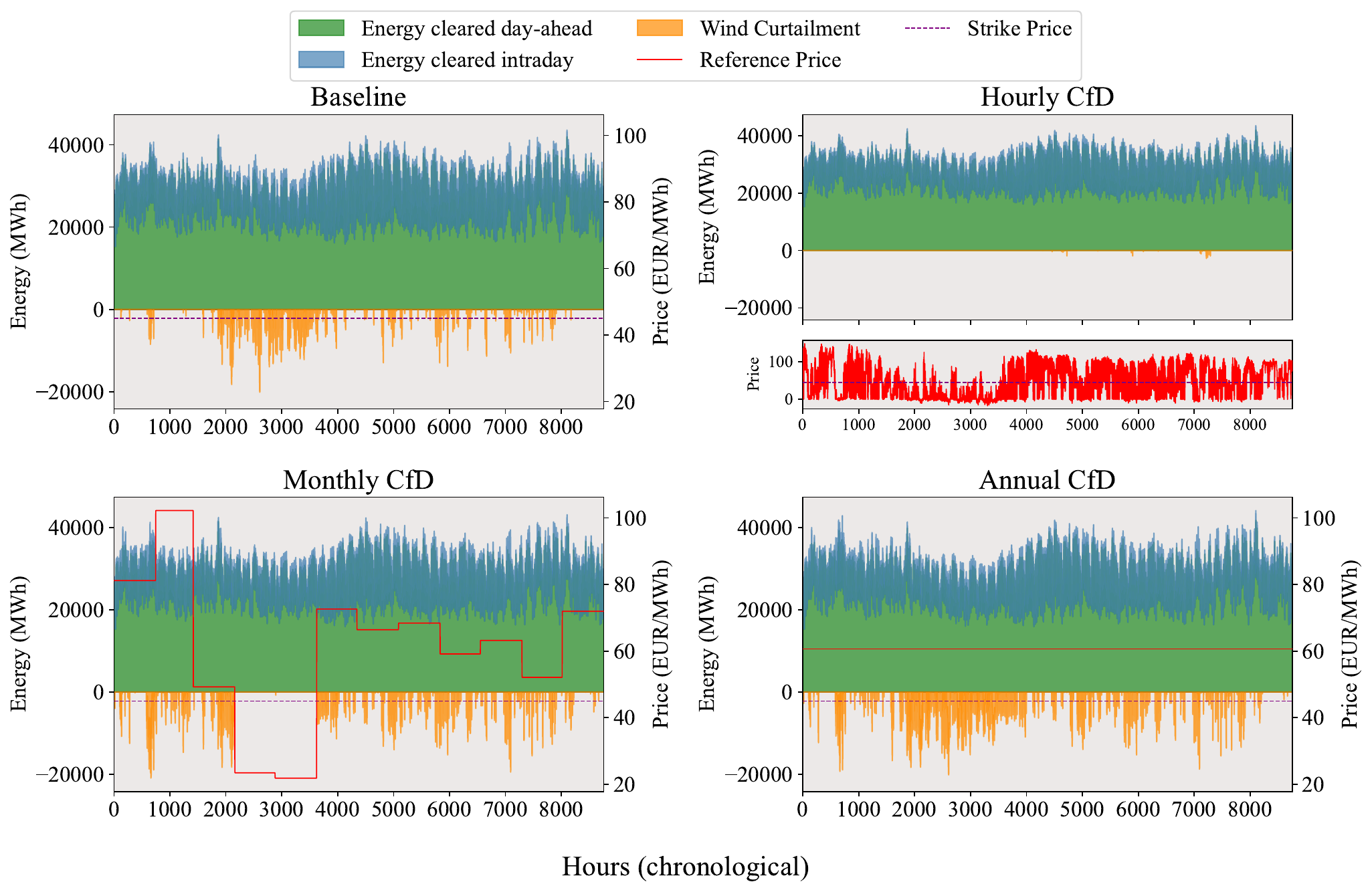}
    \caption{Hourly quantities cleared in the DA and ID markets, wind curtailment, and reference prices for the 45\,\euro{}/MWh strike price.}
    \label{fig:area_wind_curt_app}
\end{figure}

\begin{figure}[!ht]
    \centering
    \includegraphics[width=0.6\linewidth]{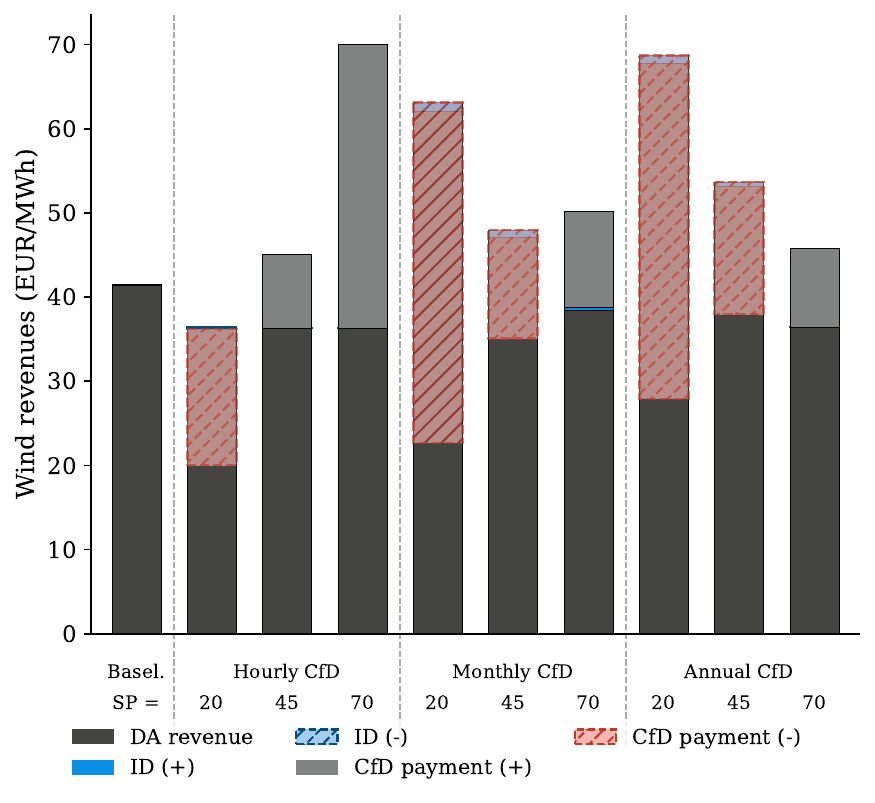}
    \caption{Wind revenues per MWh generated across the baseline and nine CfD scenarios, decomposed into market revenues and CfD settlement payments.}
    \label{fig:windprofits_app}
\end{figure}

\newpage
\begin{landscape}

\begin{figure}[!ht]
    \centering
    \includegraphics[width=1\linewidth]{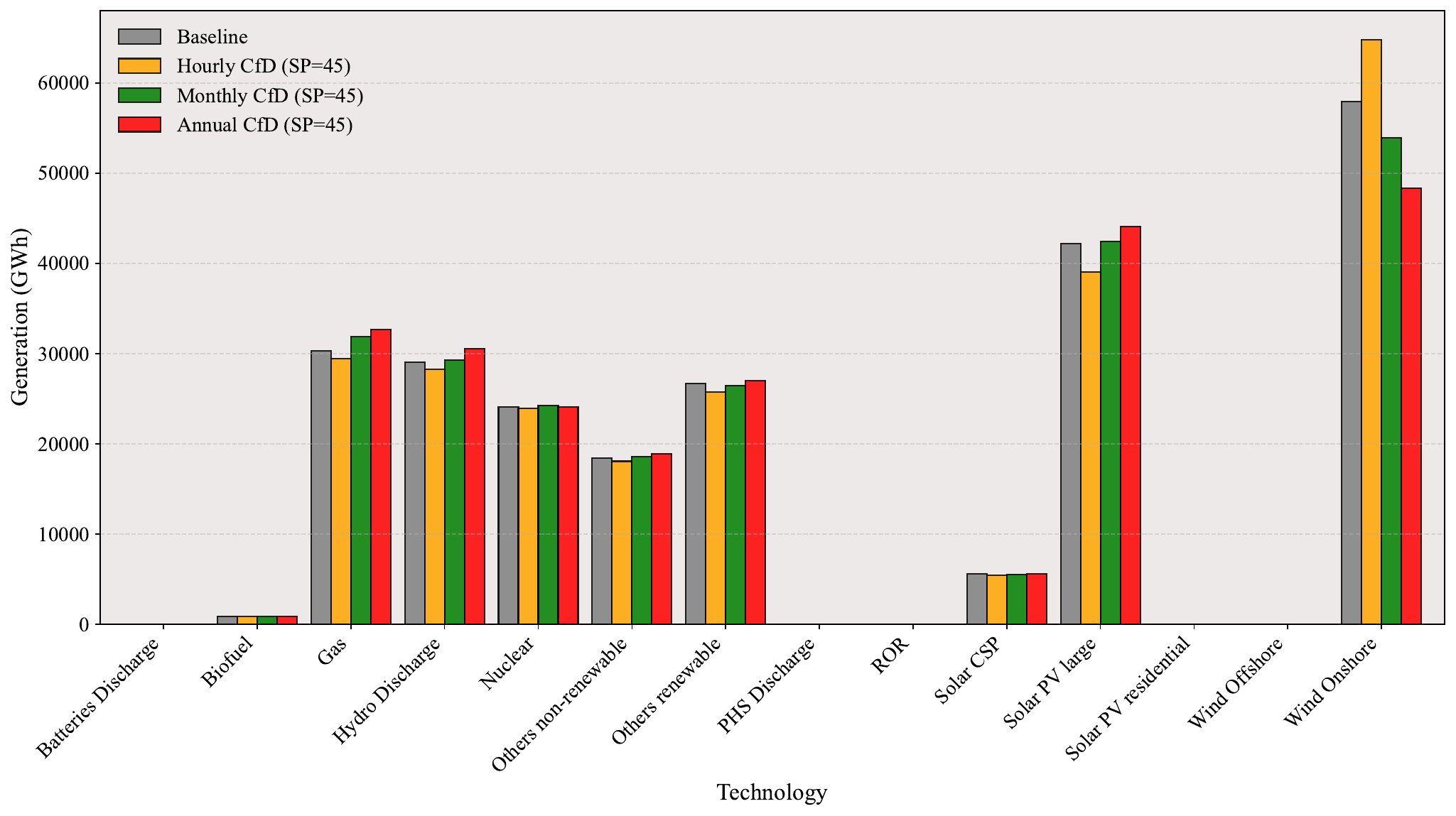}
    \caption{Total generation mix for the Baseline and CfD scenarios with a strike price of 45\,\euro{}/MWh.}
    \label{fig:generation_mix}
\end{figure}

\begin{figure}[!ht]
    \centering
    \includegraphics[width=1\linewidth]{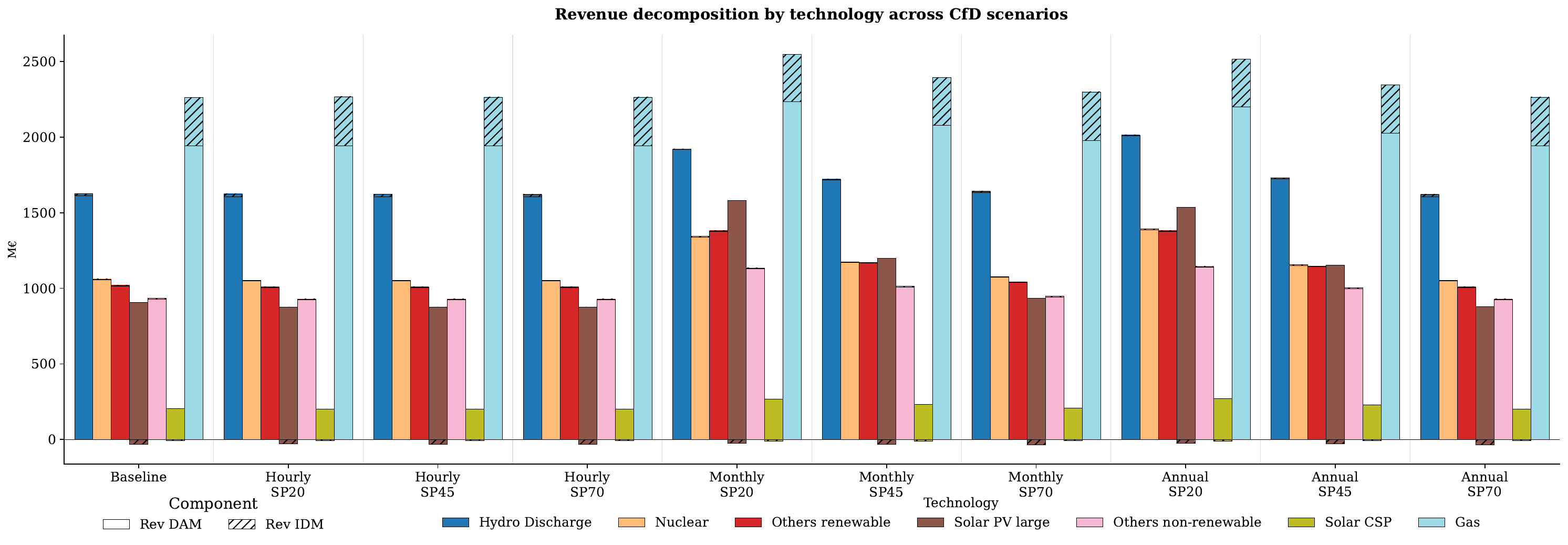}
    \caption{Profits by technology decomposed into DA and ID revenues and generation costs}
    \label{fig:profits_tech}
\end{figure}

\end{landscape}

\bibliographystyle{elsarticle-harv}
\bibliography{cas-refs}
%TC:endignore 
\end{document}